\documentclass[pdftex]{nolta}
\usepackage{amsmath} 
\usepackage{amssymb}
\usepackage{amsfonts}
\usepackage{algorithm, algorithmic}

\usepackage{textcomp}
\usepackage{bm}
\usepackage{comment}
\usepackage[whole]{bxcjkjatype} 
\usepackage{makeidx}

\usepackage[hang,small,bf]{caption}
\usepackage[subrefformat=parens]{subcaption}
\Vol{2}%
\No{3}%
\Year{2011}%
\Month{10}%

\title{Theoretical Analysis of the Three-Dimensional CAC Considering Connection and Communication Quality\\
}

\AUTHOR*{%
\author{Sota Narikiyo}{1}\orcid{0009-0009-9857-9607}, 
\author*{Sumiko Miyata}{1}\orcid{0000-0001-8023-7435},
\author*{Ken-ichi Baba}{2}\orcid{0000-0002-8170-4936},
\author*{Katsunori Yamaoka}{3}\orcid{0000-0002-1423-3337}
}

\AFFILIATE{%
\affiliate{Shibaura Institute of Technology,
3--7--5 Toyosu, Koto-ku, Tokyo, 135--8548 Japan}{1},
\affiliate{Kogakuin University,1-24-2 Nishishinjuku, shinjuku-ku, Tokyo, 163-8867 Japan}{2},
\affiliate{Tokyo Institute of Technology,2-12-1 Ookayama, Meguro-ku, Tokyo, 152-8550 Japan}{3}
}

\received{}{}{}
\revised{}{}{}
\published{}{}{}

\begin{document}

\begin{abstract}
In emergencies such as disasters, the number of voice calls (VoIP sessions) increases rapidly for a variety of purposes. Thus, a control server near a disaster area may not be able to connect to VoIP sessions due to congestion. To solve this problem, a Call Admission Control (CAC) is needed to determine whether a VoIP session requesting a connection can be accepted or rejected. A CAC has the purpose of guaranteeing the connection quality and communication quality of VoIP sessions. One conventional method classifies VoIP sessions into three classes (emergency VoIP sessions, VoIP sessions from the disaster area, and VoIP sessions from outside the disaster area) by focusing on the outgoing location and offers a CAC with a priority level for each. However, a conventional CAC method cannot be applied to VoIP networks because reception control is designed for Public Switched Telephone Networks (PSTN). When conventional methods are applied to VoIP networks, the connection quality is guaranteed, however the communication quality cannot be guaranteed because the packet dropping probability is not considered. In this paper, we propose a three-dimensional CAC that controls three classes of VoIP sessions and guarantees both communication and connection quality in VoIP networks during emergencies. A conventional CAC method and our proposed CAC method are evaluated in terms of the call blocking probability, which guarantees the connection quality, and packet dropping probability, which guarantees the communication quality, to show the effectiveness of the proposed method.

\end{abstract}



\maketitle

\section{Introduction}
In recent years, the increase in the number of base stations for mobile communications has allowed people to easily exchange information with other people in their daily lives. However, when a disaster strikes, base stations or communication lines in the disaster area get physically damaged, and victims are unable to transmit information on their own. For example, 1.9 million fixed lines were damaged in the 2011 Great East Japan Earthquake due to increased demand for safety confirmation calls \cite{tsuushinjoukyou,Mitsuyoshi}.
In other words, during a disaster, it is difficult to connect calls due to increased demand for safety confirmation calls \cite{Ran}.

 To solve this problem, which leads to traffic congestion in emergencies, Call Admission Control (CAC) is used to determine whether a call arriving in the network can be accepted \cite{yuusenntsuushin}. In an emergency, arriving calls are divided into emergency calls and general calls for safety confirmation in order to guarantee the number of accommodated emergency calls, and we call this guarantee the ``connection quality'' \cite{yuusenntsuushin}. However, CAC guarantees only the connection quality of emergency calls and does not consider the connection quality of low-priority general calls. Therefore, another CAC has been presented that guarantees the connection quality of emergency calls and allows for the maximum connection of general calls \cite{holding}. Moreover, another conventional CAC method \cite{Kawase} argues that general calls can be classified into two classes when focusing on the outgoing location. In this conventional method \cite{Kawase}, general calls are classified as general calls from the disaster area and general calls from outside the disaster area, and three classes of calls are considered: emergency calls and two classes of general calls. In addition, this conventional CAC method \cite{Kawase} is assumed to be controlled by Public Switched Telephone Networks (PSTN).

  However, approximately 60\% of emergency calls are handled by mobile phones using Internet Protocol (IP) networks \cite{NumberOfEmergencycall}. In addition, the Japanese Ministry of Internal Affairs and Communications has announced a complete transition from PSTN to IP networks by January 2025 \cite{PSNikou}. When a CAC method for PSTN is applied directly to an IP network, Quality of Service (QoS) is not guaranteed because packet dropping is not assumed. Therefore, to apply this conventional CAC method \cite{Kawase} for PSTN to IP networks, we need to consider the characteristics of packets in packet switched voice calls (VoIP sessions) using the Voice over Internet Protocol (VoIP). After a new VoIP session is allocated to the VoIP network in an emergency, new arriving packets may be dropped due to the burst arrival characteristics of voice packets because VoIP packets have bursty characteristics. The burstiness of packets must also be considered in CAC with the quality of communication guaranteed. We call this guarantee the ``communication quality.'' The conventional CAC method assumes these packet bursty characteristics by using a Markov Modulated Poisson Process (MMPP) model \cite{Narikiyo}. However, the conventional CAC method \cite{Kawase} considers only the call level characteristics and does not consider packet bursty characteristics. Thus, this CAC method guarantees only the connection quality.

In this paper, we propose a three-dimensional CAC method for emergencies that considers both the call level characteristics to evaluate the connection quality and packet level characteristics to evaluate the communication quality in VoIP networks. In this study, guaranteed communication quality is defined as a packet dropping probability that is below a set upper bound. In addition, guaranteed connection quality is defined as a call blocking probability that is below a set upper bound. First, we show that the validity of the theoretical analysis in this study is demonstrated by comparing the theoretical results with simulation results. Simulations in this paper are performed on the call level because the threshold varies significantly with the traffic load of call level. Next, the packet dropping probability and the call blocking probability of three classes of VoIP sessions considering the packet level are analyzed by varying the traffic intensity for each of the three VoIP sessions. Then, the effectiveness of our proposed method is demonstrated by comparing it with the conventional method \cite{Kawase}, which does not consider the packet level. In addition, the packet dropping probability and the call blocking probability are analyzed in more detail in order to consider our proposed method.

The following is a summary of the main contributions of this study.
\begin{itemize}
    \item Proposal of a three-dimensional CAC method for three classes of VoIP sessions considering call and packet level in an emergency. 
    \begin{itemize}
        \item Derivation of the average packet dropping probability considering the steady-state probability of the call level of the CAC method for three classes of VoIP sessions.
        \item Derivation of two optimal thresholds guaranteeing communication and connection quality.
    \end{itemize}
    \item Evaluation of connection and communication quality of the three-dimensional CAC method.
    \begin{itemize}
        \item Modeling of an 8-phase $MMPP/M/1/K$ model by queueing theory to evaluate our CAC communication quality for three classes of VoIP sessions.
        \item Demonstration of how effective our proposed CAC method is in a numerical analysis of the call blocking probability using our optimal thresholds guaranteeing the communication quality, showing that the call blocking probability of emergency VoIP sessions decreases.
    \end{itemize}
\end{itemize}

This paper is organized as follows. Section 2 explains related works on CAC methods. Section 3 explains our assumed system and our proposed CAC method. In Section 4, we model our proposed CAC method using queueing theory. Section 5 evaluates our proposed CAC method in terms of call blocking probability and packet dropping probability. Section 6 concludes this paper.

\section{Related works}

\begin{figure}[htbp]
  \begin{center}
  \hspace{-10mm}
    \includegraphics[clip,width=12.5cm,height=65mm]{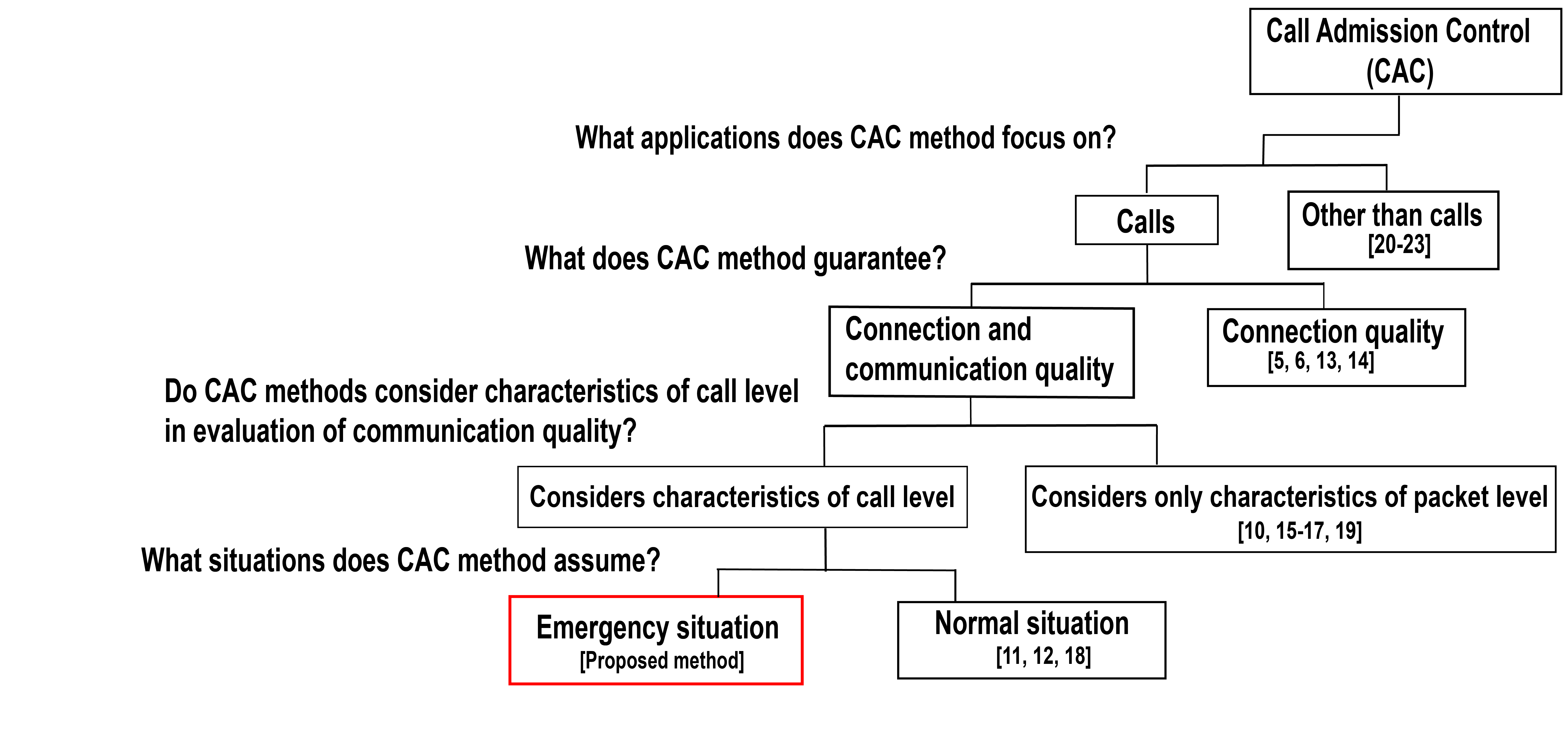}
    \caption{Classification of related works.}
    \label{references}
  \end{center}
\end{figure}

Figure \ref{references} shows the classification of related works.

Many CAC methods have been introduced for a variety of purposes. Focusing on the applications to be controlled, there are two types of CAC methods: calls \cite{Kawase,Zhou,waiting,ThreeStrategy,cognitive,Matsuoka,Jung,MaseMase,Murakami,UAV,Li,holding} and other than calls \cite{survey, cellular_CAC, static_CAC, HAP}.

A CAC method that assumes applications other than calls also cover the control of real-time applications such as video streaming and games \cite{survey, cellular_CAC, static_CAC, HAP}. However, This CAC method assumes a distributed network and not a centralized network. Therefore, CAC methods that control applications other than calls are not suitable for control servers such as a Session Initiation Protocol (SIP) server for IP networks.

Some CAC methods that assume calls are classified into guaranteed communication quality and connection quality \cite{MaseMase,UAV,ThreeStrategy,cognitive,Murakami,Matsuoka,Jung,Li} and guaranteed only connection quality \cite{holding,Kawase,Zhou,waiting}. J. Zhou et al. proposed a CAC in which the call blocking probability and preemption probability of general calls are improved by focusing on preemption and combining queueing methods to compensate for preemption deﬁcits \cite{Zhou}. This CAC method classifies calls into emergency and general calls to ensure priority for emergency calls. Another CAC method has been proposed that reduces the call blocking probability of emergency calls and general calls by waiting for general calls \cite{waiting}. However, these conventional CAC methods \cite{holding,Kawase,Zhou,waiting} guarantee only the connection quality. The evaluation of a CAC method needs to consider both the connection quality and communication quality of VoIP sessions.

 Some CAC methods for evaluating the two types of quality can be further classified into two categories. One type of CAC method evaluates communication quality by considering call level characteristics with the packet dropping probability \cite{ThreeStrategy,cognitive,Murakami}. Another type of CAC method evaluates communication quality by considering only packet level characteristics with the packet dropping probability  \cite{MaseMase,UAV,Matsuoka,Jung,Li}. K. Mase proposed a CAC method that achieves high link utilization and satisfies the QoS of all users in a VoIP network \cite{MaseMase}. Moreover, another conventional method \cite{UAV}, a CAC for VoIP, was proposed for WiFi access networks deployed by an Unmanned Aerial Vehicle (UAV) relaying to a 5G network. However, these CAC methods \cite{MaseMase,UAV,Matsuoka,Jung,Li} are not suitable for evaluating communication quality because the call level characteristics are not considered in the derivation of the packet dropping probability.

Among the CAC methods that consider the call level characteristics applied to a packet dropping probability, the assumed situation can be further classified into emergency situations and normal situations \cite{ThreeStrategy,cognitive,Murakami}. Castellanos-Lopez et al. proposed a CAC method from three perspectives: user level, active VoIP session, and packet level \cite{ThreeStrategy}. Moreover, they applied a previous CAC \cite{ThreeStrategy} to cognitive radio networks and proposed a method with three strategies with priorities \cite{cognitive}. R. Murakami et al. proposed a CAC method for determining the maximum number of accommodated VoIP sessions using the average packet dropping probability in consideration of the state probability \cite{Murakami}. However, these conventional CAC methods \cite{ThreeStrategy,cognitive,Murakami} are difficult to apply to emergency situations where the traffic load increases rapidly and VoIP sessions need to be prioritized. 

Based on the classification of related works, the novelty of our proposed CAC method is shown in Table \ref{ComparisonResult}.
\begin{table}[H]
    \caption{Comparison of related works.}
    \begin{center}
        \scalebox{0.85}{
        \begin{tabular}{|l|r|r|r|r|} \hline
                &  
            \begin{tabular}{c}
                A
            \end{tabular}
                & 
            \begin{tabular}{c}
                B
            \end{tabular}   
                & 
            \begin{tabular}{c}
                C
            \end{tabular} 
                &
            \begin{tabular}{c}
                D
            \end{tabular}\\ \hline
            
            \begin{tabular}{c}
                \cite{UAV}
            \end{tabular} & ✓  &  &  & \\ \hline
            
            \begin{tabular}{c}
                \cite{ThreeStrategy, cognitive}
            \end{tabular} & ✓ & ✓ &  & \\ \hline
            
            \begin{tabular}{c}
                \cite{Kawase}
            \end{tabular} &  &  & ✓ & ✓\\ \hline
            \begin{tabular}{c}
                The proposed CAC method
            \end{tabular}  &✓  &✓  &✓  &✓ \\ \hline
        \end{tabular}
        }
    \end{center}
    \vspace{3mm}
    \scalebox{0.85}{A: The CAC method guarantees both the connection and communication quality.}
    \\\scalebox{0.85}{B: The CAC methods consider characteristics of call level in evaluation of communication quality.}
    \\\scalebox{0.85}{C: The CAC methods assume an emergency situation such as a disaster.}
    \\\scalebox{0.85}{D: The CAC method controls three classes of calls.}
    \label{ComparisonResult}
\end{table}

\section{Proposed method}
\subsection{Assumed system}

\begin{figure}[ht]
  \begin{center}
    \includegraphics[clip,width=10cm,height=60.0mm]{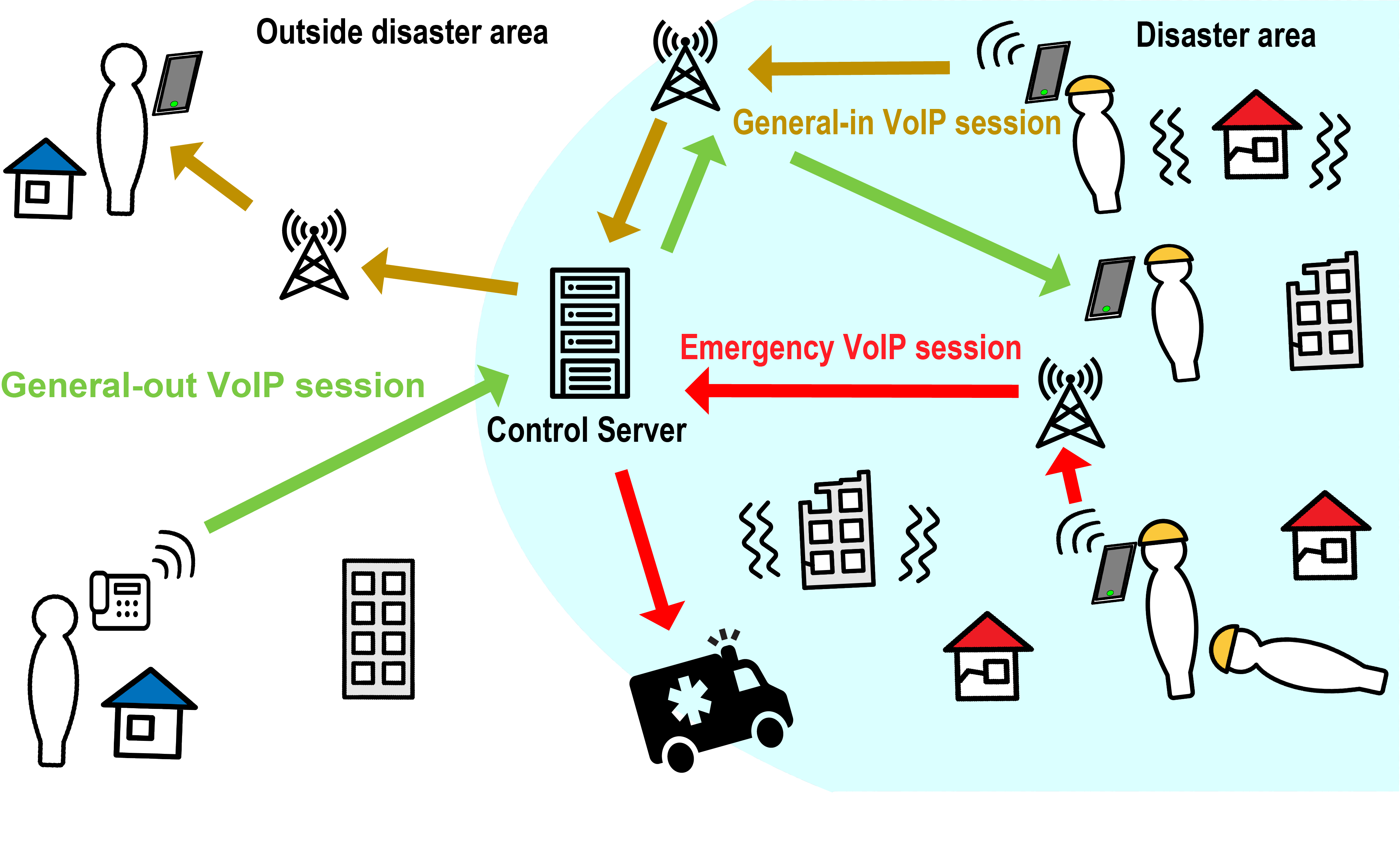}
    \caption{Assumed environment.}
    \label{overview_system}
  \end{center}
\end{figure}

To propose a novel CAC method for emergencies, this study assumes VoIP sessions that can be connected to emergency calls. As shown in Fig. \ref{overview_system}, we assume that all VoIP sessions are controlled by a control server, such as an SIP server. A VoIP session where the destination is to an emergency disaster area such as an ambulance service (red arrow) is defined as an ``emergency VoIP session.'' A VoIP session from the disaster area (yellow arrows) is defined as a ``general-in VoIP session.'' A VoIP session from outside the disaster area (green arrows) is defined as a ``general-out VoIP session.'' Therefore, each VoIP session that arrives at the control server is classified by focusing on the outgoing location. First, VoIP sessions that arrive at the control server are classified as emergency calls and general calls, and general calls are then further classified by connection location into general-in calls and general-out calls.

\subsection{Proposed CAC and objective function}
\begin{figure}[htb]
    \centering
    \hspace{25mm}
    \includegraphics[clip,width=12.5cm,height=65mm]{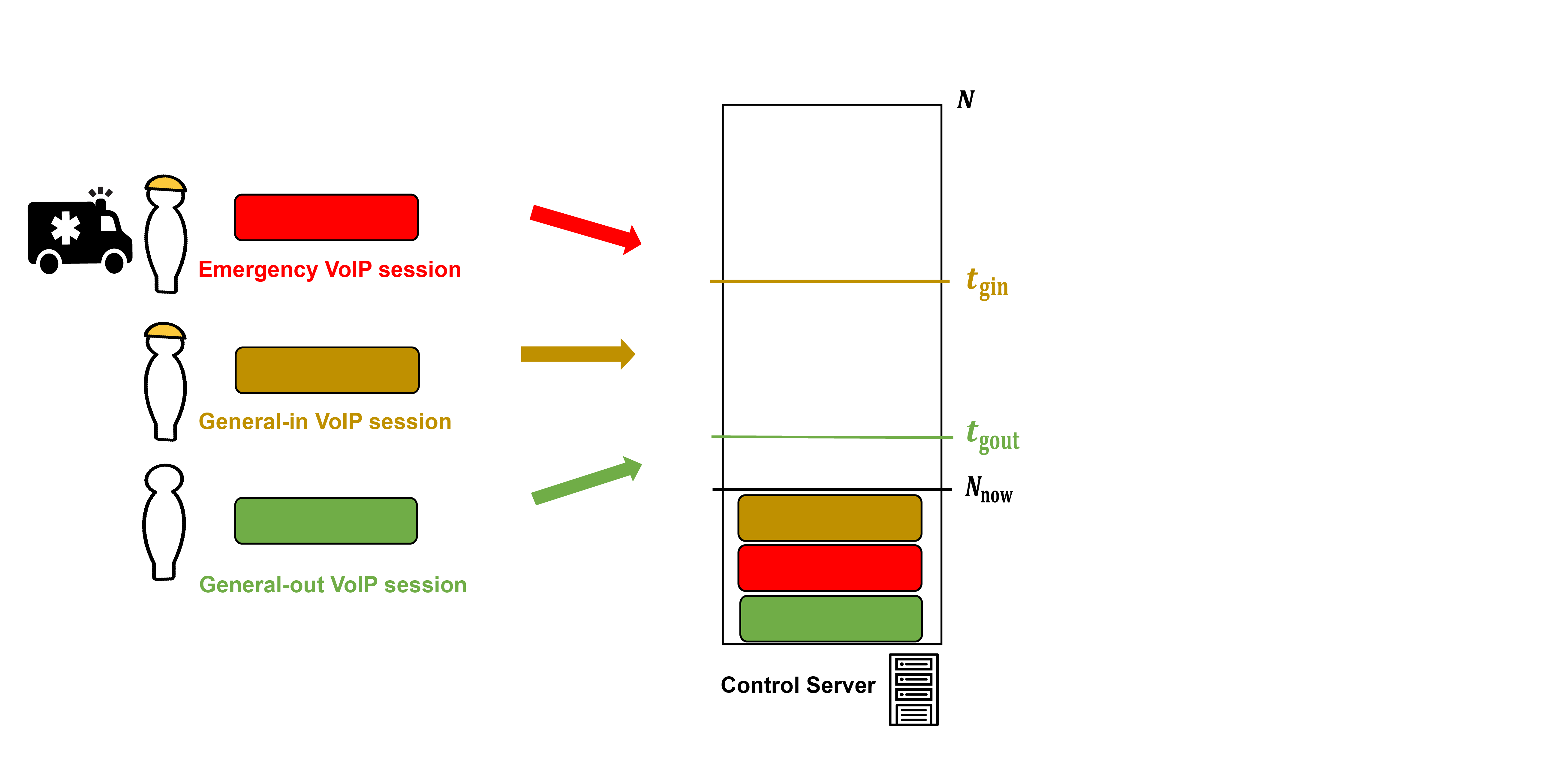}
    \caption{Overview of threshold control when not all VoIP sessions are call blocked.}
    \label{Nocontrol}
\end{figure}

\begin{figure}[htb]
    \centering
    \hspace{25mm}
    \includegraphics[clip,width=12.5cm,height=65mm]{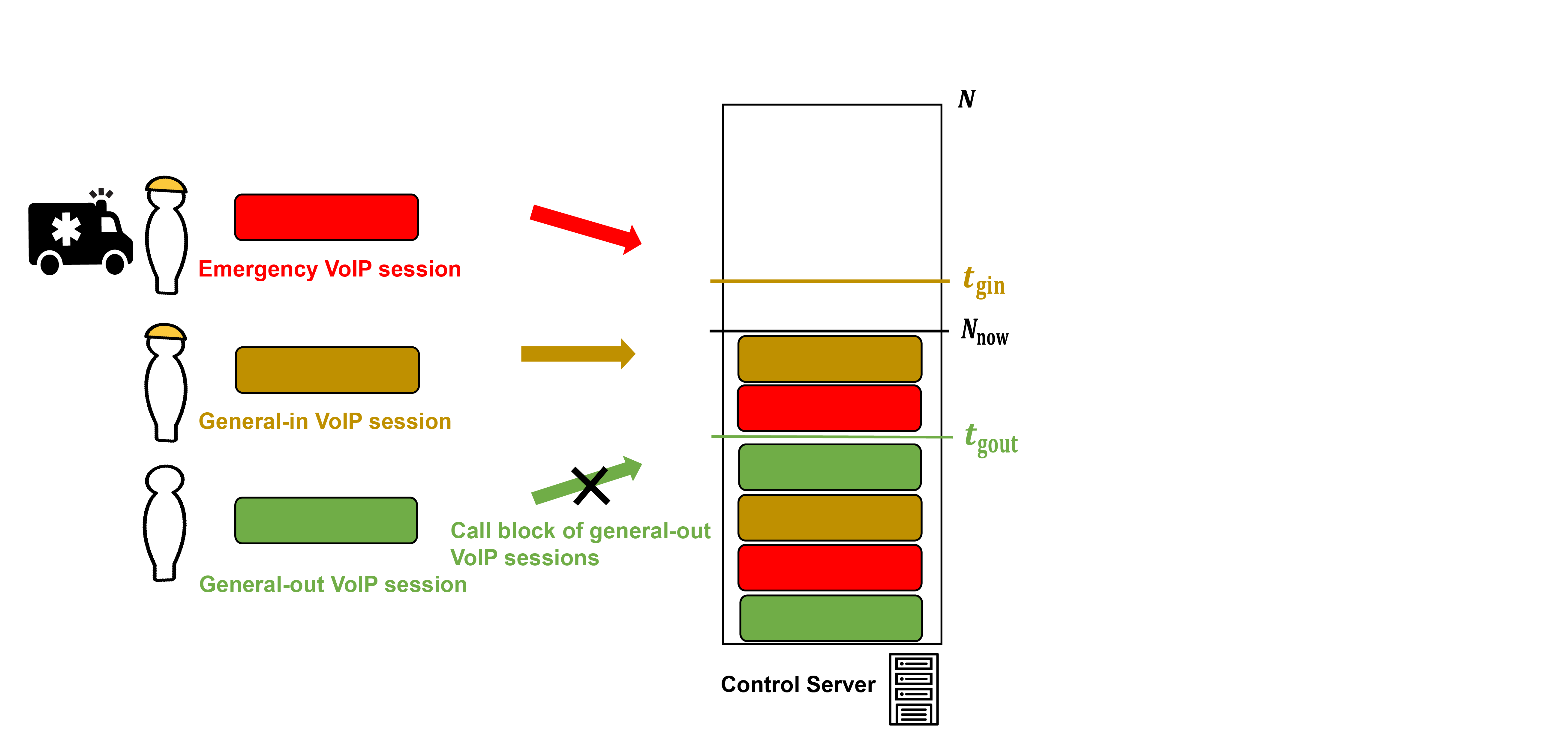}
    \caption{Overview of threshold control when general-out VoIP sessions are call blocked.}
    \label{rgout}
\end{figure}

\begin{figure}[htb]
    \centering
    \hspace{25mm}
    \includegraphics[clip,width=12.5cm,height=65mm]{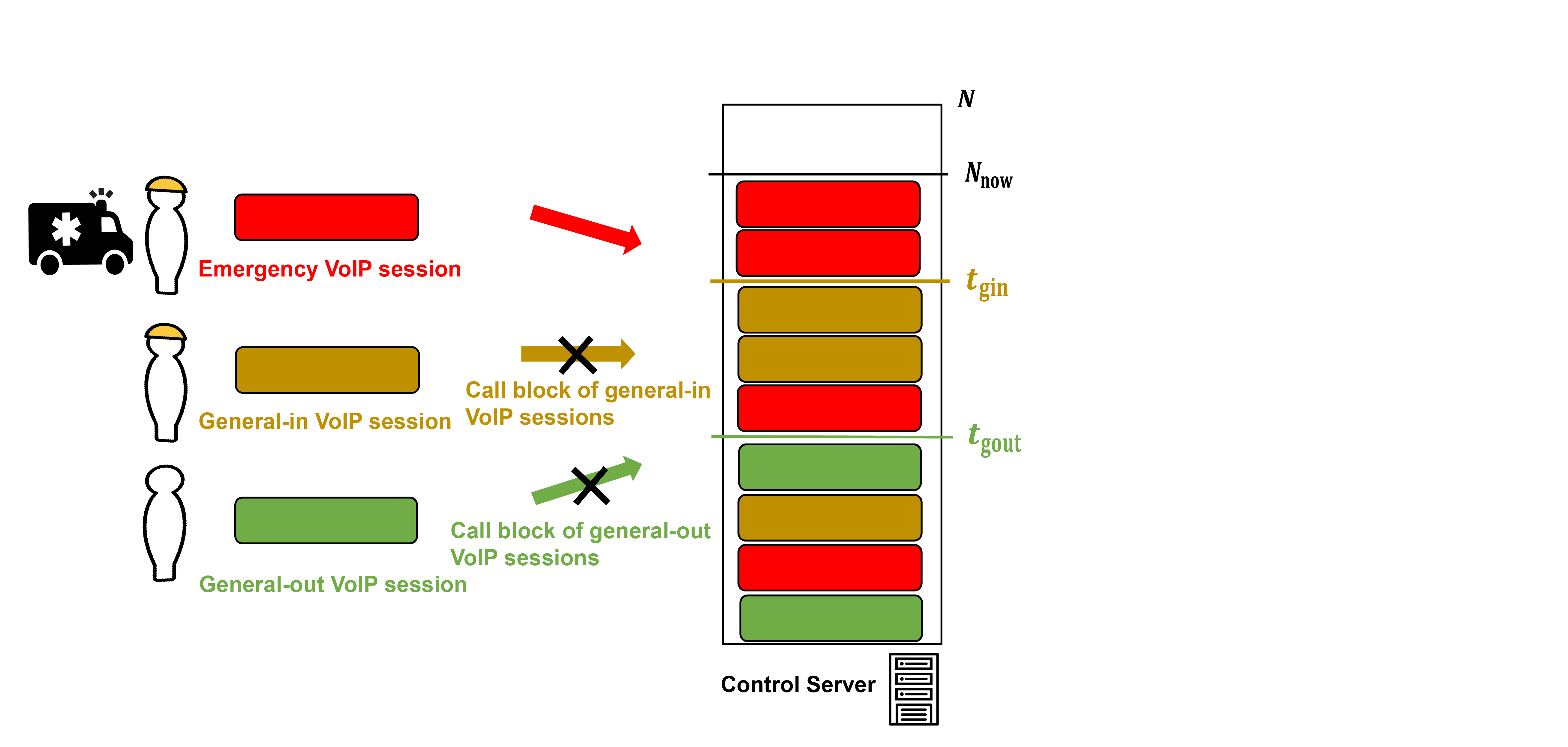}
    \caption{Overview of threshold control when general-in VoIP sessions are call blocked.}
    \label{rgin}
\end{figure}

In this paper, we analyze the connection quality and communication quality when the CAC method with the three classes of calls of the conventional method \cite{Kawase} is applied to a VoIP network. Our proposed CAC method classifies each VoIP session into three classes: emergency VoIP sessions, general-in VoIP sessions, and general-out VoIP sessions. The number of emergency VoIP sessions accommodated by the control server is $n_{\mathrm{e}}$ [number], the number of general-in VoIP sessions accommodated is $n_{\mathrm{gin}}$ [number], and the number of general-out VoIP sessions accommodated is $n_{\mathrm{gout}}$ [number]. In addition, let $\mathcal N$ be the set of all combinations of VoIP sessions. Our proposed CAC method sets a priority level for each VoIP session and selects the VoIP session to be blocked in accordance with the combination vector ${\bm n}_a =(n_{\mathrm{e}}, n_{\mathrm{gin}}, n_{\mathrm{gout}}), {\bm n}_a \in \mathcal N$ of the VoIP sessions accommodated $(n_{\mathrm{e}}, n_{\mathrm{gin}}, n_{\mathrm{gout}})$ at that time.

Each VoIP session has a priority level for every class; the highest priority level is the emergency VoIP session, the middle priority level is the general-in VoIP session, and the lowest priority level is the general-out VoIP session. In addition, each threshold $(t_{\mathrm{gin}}$, $t_{\mathrm{gout}})$ for the general-in and general-out VoIP sessions are control parameters that reserve bandwidth for arrival VoIP sessions of higher priority than each threshold. Let $\bm t$ be a vector with each threshold $(t_{\mathrm{gin}}$, $t_{\mathrm{gout}})$ for the general-in and general-out VoIP sessions. Let ${\mathcal N}_{\bm t}$,  $1\leqq i \leqq |{\mathcal N}_{\bm t}|$ be the total number of combinations of each VoIP session up to a threshold vector $\bm t$.

From Figs. \ref{Nocontrol}--\ref{rgin}, our proposed CAC method at the control server compares each threshold $(t_{\mathrm{gin}}$, $t_{\mathrm{gout}})$ and number of accommodated VoIP sessions $N_{\mathrm{now}}$. In addition, the maximum number of accommodated VoIP sessions $N$ in the link bandwidth is used. First, from Fig. \ref{Nocontrol}, when $N_{\mathrm{now}}$ is not over $t_{\mathrm{gout}}$, each arriving VoIP session is accommodated at the control server. Second, from Fig. \ref{rgout}, when $N_{\mathrm{now}}$ is over $t_{\mathrm{gout}}$, the control server blocks general-out VoIP sessions. Moreover, from Fig. \ref{rgin}, when $N_{\mathrm{now}}$ is over $t_{\mathrm{gin}}$, the control server blocks general-in VoIP sessions. Finally, when $N_{\mathrm{now}}$ is over $N$, the control server blocks emergency VoIP sessions. In other words, emergency VoIP sessions are blocked, which means that all three classes of VoIP sessions are blocked.

The probability of each VoIP session being blocked is expressed as the ``call blocking probability,'' which is an evaluation value of the connection quality. The probability of packets being dropped in the control server is expressed as the ``packet dropping probability'' which is an evaluation value of the communication quality.

When each VoIP session arrives, the acceptance or non-acceptance of the VoIP sessions is determined by a threshold. Let $L^{\mathrm{b}}_{\mathrm{e}}$ be the call blocking probability of emergency VoIP sessions, $L^{\mathrm{b}}_{\mathrm{gin}}$ be the call blocking probability of general-in VoIP sessions, $L^{\mathrm{b}}_{\mathrm{gout}}$ be the call blocking probability of general-out VoIP sessions, and $C_{\mathrm{e}}$ and $C_{\mathrm{gin}}$ be upper bound call blocking probabilities for emergency VoIP sessions and general-in VoIP sessions, respectively. In addition, let $\bm t^*:=(t^*_{\mathrm{gin}},t^*_{\mathrm{gout}})$ be the optimal threshold vector. Let $\Bar{L}^{\mathrm{d}}({\mathcal N}_{\bm t})$ be the average packet dropping probability, and $\Bar{L}^{\mathrm{d}}_{\mathrm{upper}}$ be the upper bound average packet dropping probability.

 Based on the above, our objective function of this study is 
 
\begin{align*}
\bm{t}^*=(t^*_{\mathrm{gin}},t^*_{\mathrm{gout}})=\underset{\bm t}{\mathrm{argmin}}\quad L^{\mathrm{b}}_{\mathrm{gout}}, \notag\\
\mathrm{s.t}\quad L^{\mathrm{b}}_{\mathrm{e}}\leqq C_{\mathrm{e}}, L^{\mathrm{b}}_{\mathrm{gin}}\leqq C_{\mathrm{e}}, C_{\mathrm{e}}\leqq C_{\mathrm{gin}},\\L^{\mathrm{b}}_{\mathrm{gin}}\leq L^{\mathrm{b}}_{\mathrm{gout}}, \Bar{L}^{\mathrm{d}}({\mathcal N}_{\bm t})\leqq \Bar{L}^{\mathrm{d}}_{\mathrm{upper}},
\end{align*}
where the condition is not satisfied, each threshold $t_{\mathrm{gin}}$ and $t_{\mathrm{gout}}$ is set to 0, and only emergency VoIP sessions are accepted.

In this study, we analyze all combination patterns of thresholds to derive the optimal threshold $(t^*_{\mathrm{gin}}, t^*_{\mathrm{gout}})$ that satisfies the objective function. Moreover, we derive the call blocking probability of each VoIP session and the packet dropping probability in the control server. From the results, we derive $(t^*_{\mathrm{gin}}, t^*_{\mathrm{gout}})$, which adds a guarantee of communication quality to the conventional method \cite{Kawase}.

\section{Modeling}

\subsection{Modeling by queueing theory}

\begin{figure}[htb]
    \centering
    \includegraphics[clip,width=12cm,height=65.0mm]{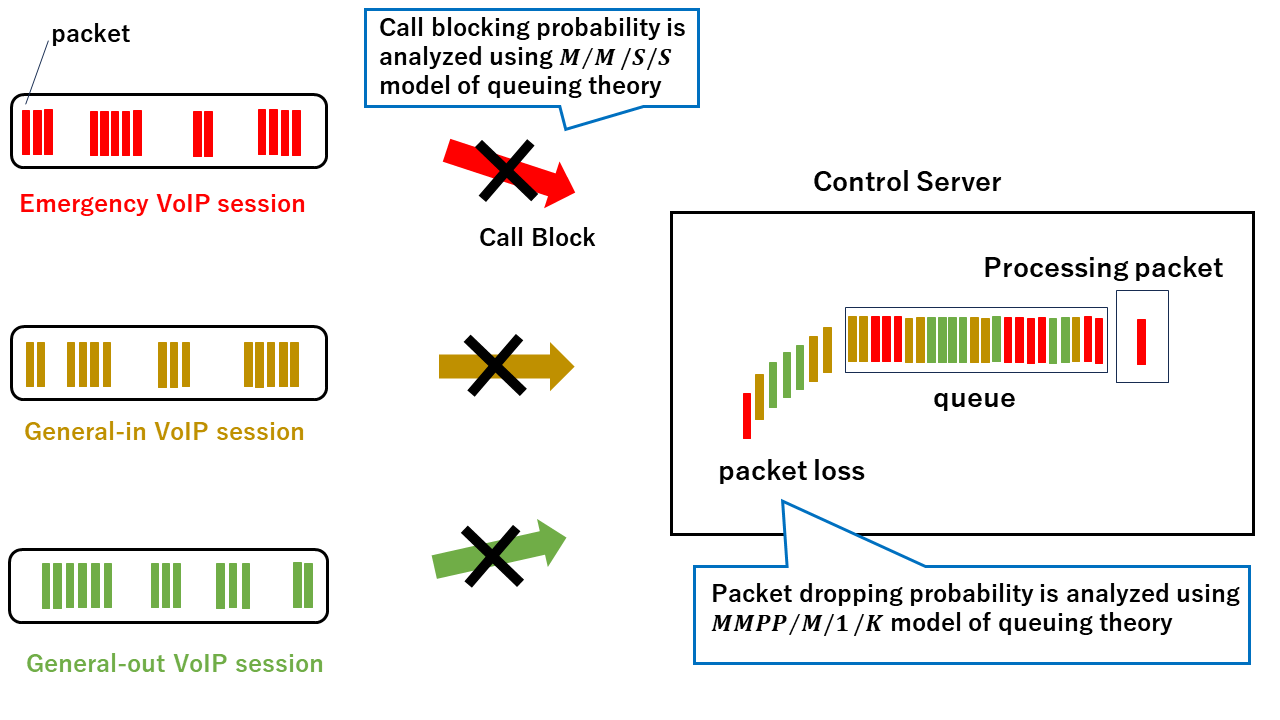}
    \caption{Overview of threshold control.}
    \label{packet}
\end{figure}

\begin{figure}[ht]
    \hspace{-2.5mm}
    \begin{minipage}[t]{0.5\linewidth}
        \centering
        \includegraphics[scale=0.07]{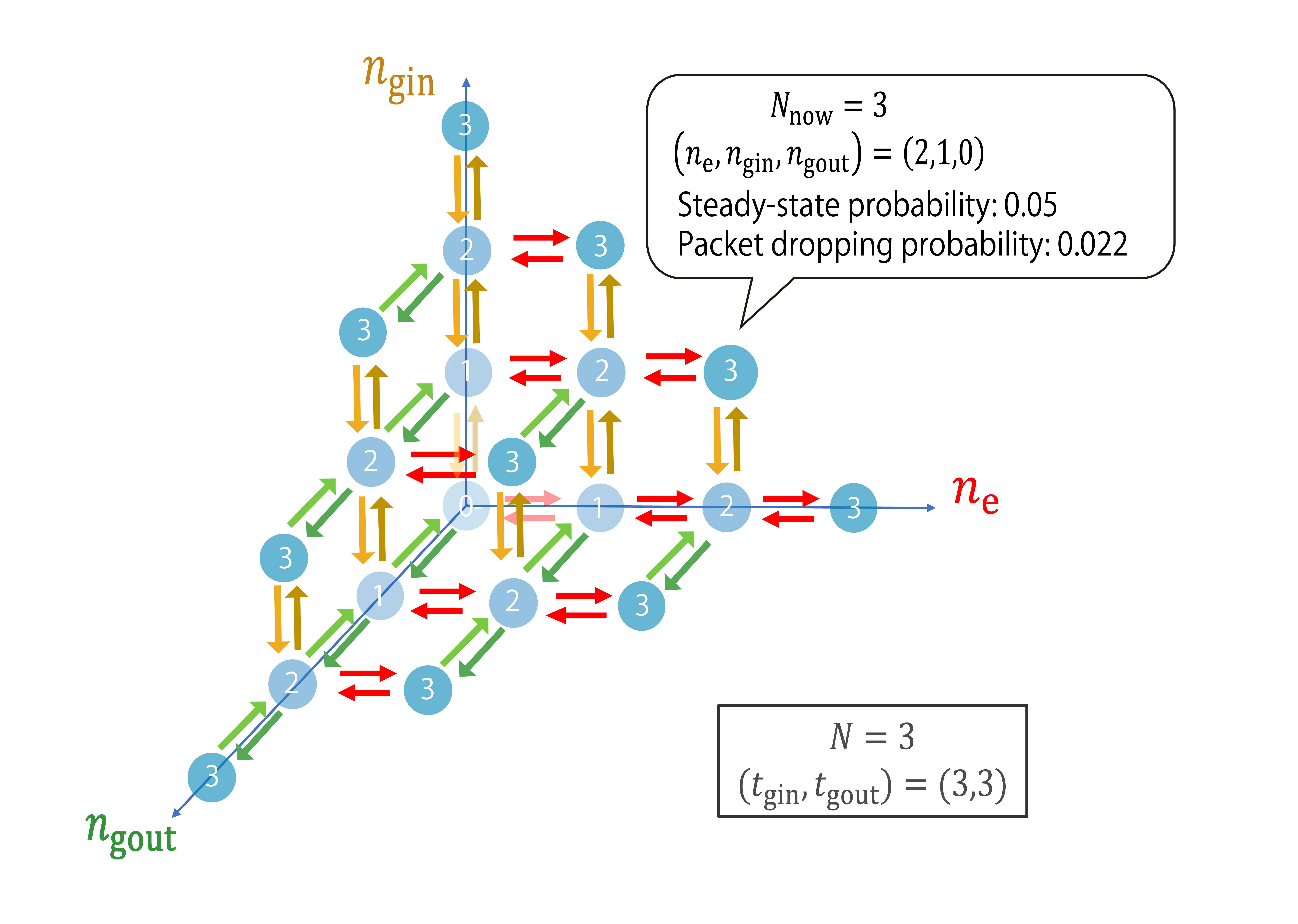}
        \subcaption{Steady-state space when each threshold $(t_{\mathrm{gin}}, t_{\mathrm{gout}})$ is (3, 3).}
        \label{StateSpace_33}
    \end{minipage}
    \begin{minipage}[t]{0.5\linewidth}
        \centering
        \includegraphics[scale=0.07]{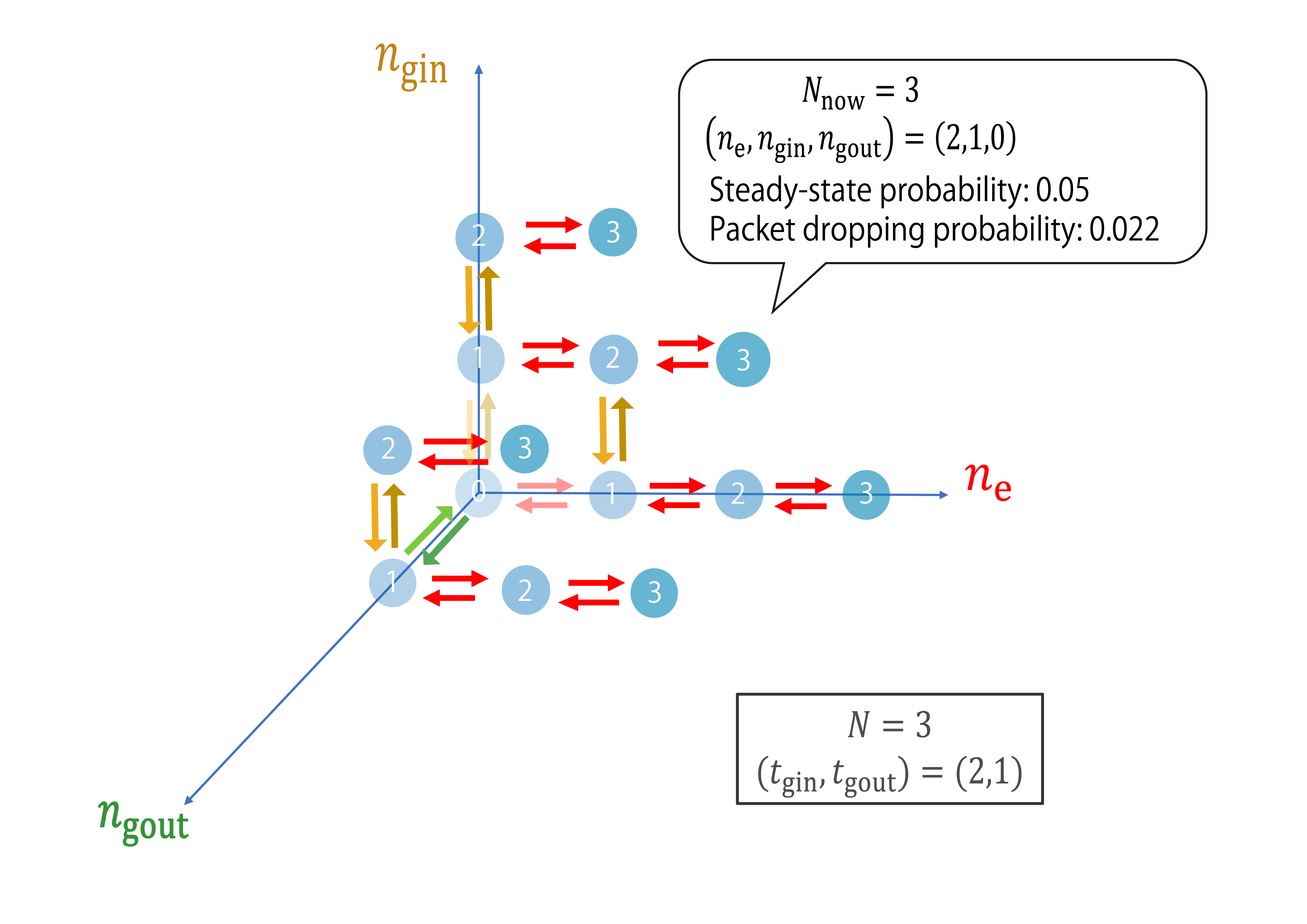}
        \subcaption{Steady-state space when each threshold $(t_{\mathrm{gin}}, t_{\mathrm{gout}})$ is (2, 1).}
        \label{StateSpace_21}
    \end{minipage}
    \caption{Steady-state space of call level.}
    \label{StateSpace_flow}
\end{figure}

In this paper, our proposed CAC method is analyzed from two aspects: connection quality by using call blocking probability and communication quality by using packet dropping probability. For each analysis, different models of queueing theory are treated. From Fig. \ref{packet}, the call blocking probability for the connection quality is derived by using the $M/M/s/s$ model of queueing theory. In the $M/M/s/s$ model, customers are assumed to be VoIP sessions, and the server is assumed to be a control server that is expected to be congested in an emergency. On the other hand, the packet dropping probability for communication quality is derived by using the $MMPP/M/1/K$ model of queueing theory. In the $MMPP/M/1/K$ model, customers are assumed to be voice packets, and the server is assumed to be the same control server as in the $M/M/s/s$ model. In addition, the analysis with the $MMPP/M/1/K$ model also needs to consider the steady-state probability of the call level, which is affected by the traffic parameters and thresholds of the VoIP session.

 Fig. \ref{StateSpace_flow} shows the steady-state space of VoIP sessions. Each axis is the number of each VoIP session accommodated $(n_{\mathrm{e}}, n_{\mathrm{gin}}, n_{\mathrm{gout}})$. The number in the blue circle means the total number of VoIP sessions $N_{\mathrm{now}}$. The arrows represent transitions (arrivals and departures) of each VoIP session. In Figs. \ref{StateSpace_33} and \ref{StateSpace_21}, the maximum number of accommodated VoIP sessions $N$ is set to 3. In Fig. \ref{StateSpace_33}, each threshold is assumed to be $(t_{\mathrm{gin}}, t_{\mathrm{gout}})= (3,3)$, while in Fig. \ref{StateSpace_21}, each threshold is assumed to be $(t_{\mathrm{gin}}, t_{\mathrm{gout}})= (2,1)$. Therefore, the number of states in Fig. \ref{StateSpace_33} decreases from the number of states in Fig. \ref{StateSpace_21} because the thresholds in Fig. \ref{StateSpace_21} are smaller than in Fig. \ref{StateSpace_33}. 
 
 One blue circle has the steady-state probability of the call level, and the packet dropping probability exists in each VoIP session combination ${\bm n}_a$. The steady-state probability of the call level varies with the traffic intensity of each VoIP session. This is because transition probabilities such as arrival and departure rates depend on traffic intensity. In addition, from Figs. \ref{StateSpace_33} to \ref{StateSpace_21}, $(t_{\mathrm{gin}}, t_{\mathrm{gout}})$ controls the number of the transition probability and changes the number of each VoIP session combination ${\bm n}_a$. Thus, the steady-state probability changes with $(t_{\mathrm{gin}}, t_{\mathrm{gout}})$. On the other hand, changing $(t_{\mathrm{gin}}, t_{\mathrm{gout}})$ does not change the packet dropping probability. This is because, as shown in the numerical result in Figs. \ref{StateSpace_33} to \ref{StateSpace_21}, the packet dropping probability depends only on the combination of each VoIP session ${\bm n}_a$. 
 
 Thus, to evaluate communication quality, the steady-state probability of the call level must be considered in the packet dropping probability using the $MMPP/M/1/K$ model. In this study, the average packet dropping probability $\Bar{L}^{\mathrm{d}}({\mathcal N}_{\bm t})$ is determined in order to evaluate the communication quality using the steady-state probability of the call level derived by the $M/M/s/s$ model and the packet dropping probability derived by the $MMPP/M/1/K$ model. Table 1 shows the notation for each parameter when modelling using the $M/M/s/s$ model of queueing theory.

\begin{table}[H]
    \caption{Notations of M/M/s/s model.}
    \centering
    \scalebox{0.75}{
    \begin{tabular}{c||c}
    \hline \hline
    Notation & Description \\
    \hline \hline
    $n_{\mathrm{e}}$, $n_{\mathrm{gin}}$, $n_{\mathrm{gout}}$  & Number of each VoIP session accommodated\\ \hline
    ${\bm n}_a$ &  Vector of each VoIP session combination\\ \hline
    $\mathcal N$ &  Set of ${\bm n}_a$\\ \hline
    $t_{\mathrm{gin}}$ & Threshold for general-in VoIP session\\ \hline
    $t_{\mathrm{gout}}$ &  Threshold for general-out VoIP session\\ \hline
    $\bm{t}$ &  Vector of threshold \\ \hline
    ${\mathcal N}_{\bm t}$ & Set of ${\bm n}_a$ considering threshold \\ \hline
    $t^*_{\mathrm{gin}}$ &  Optimal threshold for general-in VoIP session \\ \hline
    $t^*_{\mathrm{gout}}$ &   Optimal threshold for general-out VoIP session \\ \hline
    $\bm{t}^*$ &  Vector of optimal threshold\\ \hline
    $P^{\mathrm{session}}_{{\bm n}_a}$ &  Steady-state probability for each VoIP session combination\\ \hline
    $\lambda^{\mathrm{session}}_{\mathrm{e}}$, $\lambda^{\mathrm{session}}_{\mathrm{gin}}$, $\lambda^{\mathrm{session}}_{\mathrm{gout}}$ &  Arrival rate of each VoIP session\\ \hline
    $\mu^{\mathrm{session}}_{\mathrm{e}}$, $\mu^{\mathrm{session}}_{\mathrm{gin}}$, $\mu^{\mathrm{session}}_{\mathrm{gout}}$&  Departure rate of each VoIP session\\ \hline
    
    $\rho_{\mathrm{e}}$, $\rho_{\mathrm{gin}}$, $\rho_{\mathrm{gout}}$&  Traffic intensity of each VoIP session\\ \hline
    $N$ &  Maximum number of accommodated VoIP sessions\\ \hline
    $L^{\mathrm{b}}_{\mathrm{e}}$, $L^{\mathrm{b}}_{\mathrm{gin}}$, $L^{\mathrm{b}}_{\mathrm{gout}}$ & Call blocking probability of each VoIP session \\ \hline
    $C_{\mathrm{e}}$ &  Upper bound call blocking probability for emergency VoIP session \\ \hline
    $C_{\mathrm{gin}}$ &  Upper bound call blocking probability for general-in VoIP session \\ \hline
    $N_{\mathrm{now}}$ &  Total number of VoIP sessions \\ \hline 
    $\mathcal D$ &  
    \begin{tabular}{c} 
    Set of all possible combinations of VoIP sessions that can exist \\in all situations A, B, and C
    \end{tabular}\\ \hline
    \hline
    \end{tabular}
    }
    \label{MMSSpara}
\end{table}

Next, Table \ref{MMPPpara} shows the notation for each parameter when modelling using the $MMPP/M/1/K$ model of queueing theory.
\begin{table}[H]
    \caption{MMPP/M/1/K model notation.}
    \centering
    \scalebox{0.75}{
    \begin{tabular}{c||c}
    \hline \hline
    Notation & Description \\
    \hline \hline
    $\Bar{L}^{\mathrm{d}}({\mathcal N}_{\bm t})$ & Average packet dropping probability\\ \hline
    $\Bar{L}^{\mathrm{d}}_{\mathrm{upper}}$ &  Upper bound average packet dropping probability\\ \hline
    $\alpha^{-1}$ &  Talkspurt period\\ \hline
    $\beta^{-1}$ &  Silent period\\ \hline
    $T$ & Packet arrival interval\\ \hline
    $\lambda_{\mathrm{p}}$ & Arrival rate of packets \\ \hline
    $C^{2}{\mathrm{a}}$ &  Square variation coefficient\\ \hline
    $S{\rm k}$ &  Skewness\\ \hline
    $n$ &  Number of accommodated VoIP sessions for one class \\ \hline
    $r^{-1}_{0}, r^{-1}_{1}$ &  Packet mean inter-arrival time per phase (0: dense, 1: sparse)\\ \hline
    $\lambda^{\mathrm{packet}}_{0}, \lambda^{\mathrm{packet}}_{1}$ &  Arrival rate of packets per phase \\ \hline
    $q_{0}, q_{1}$ & Transition rate of phase\\ \hline
    $Q_{\mathrm{single}}$ &  
    \begin{tabular}{c} 
        Infinitesimal generating operators of continuous-time Markov chains \\for a single VoIP sessions
    \end{tabular}\\ \hline
    $\Lambda^{\mathrm{packet}}_{\mathrm{single}}$ &  Arrival rate in each phase of a single VoIP session\\ \hline
    $i$ & Number of VoIP session classes\\ \hline
    $c$ &  Total number of VoIP session classes\\ \hline
    $Q_i(Q)$ & 
    \begin{tabular}{c} 
        Infinitesimal generating operators of continuous-time Markov chains \\ for VoIP sessions of $i(i=c)$ classes
    \end{tabular}\\ \hline
    $\Lambda_i^{\mathrm{packet}}(\Lambda^{\mathrm{packet}})$ &  Arrival rate for each phase of VoIP sessions of $i(i=c)$ classes\\ \hline
    $\bm{M}_i(\bm{M})$ &  Parameters representing MMPP for VoIP sessions of $i(i=c)$ classes\\ \hline
    $q_{ijk}$ & Three classes of VoIP session phases\\ \hline
    $\lambda^{\mathrm{packet}}_{ijk}$ &  Packet arrival rate of three classes of VoIP sessions\\ \hline
    $K$ & Capacity of queue \\ \hline
    $m$ &  Number of packets waiting in queue\\ \hline
    
    $P^{\mathrm{packet}}_{ijk}(m)$ &  
    \begin{tabular}{c}
        Steady-state probability when number of packets waiting \\in queue is $m$
    \end{tabular}\\ \hline
    
    $a_m$ &  
    \begin{tabular}{c}
        Control variables in the packet level state equation
    \end{tabular}\\ \hline
    
    $\mathcal M$ & 
    \begin{tabular}{c}
        State space of combinatorial vector $(i,j,k,m)$ \\representing states of 8 phases
    \end{tabular}\\ \hline
    
    ${\mathcal S}$ &  Space of each phase $(i,j,k)$ \\ \hline
    $L^{\mathrm{d}}({\bm n}_a)$ &  Packet dropping probability per VoIP session combination\\ \hline
    $B$ & Bandwidth\\ \hline
    $z$ & Mean packet size\\ \hline
    $\mu^{\mathrm{packet}}$ & Packet processing rate\\ \hline
    \hline
    \end{tabular}
    }
    \label{MMPPpara}
\end{table}

Our proposed CAC method assumes three classes of VoIP sessions in emergency. Therefore, our method can be expressed as the $M_1M_2M_3/M_1M_2M_3/S/S$ model in queueing theory. The steady-state probability $P^{\mathrm{session}}_{{\bm n}_a}$ is the state probability when the state in the control server is ${\bm n}_a$. The mean arrival rates of VoIP sessions for emergency VoIP sessions, general-in VoIP sessions, and general-out VoIP sessions follow a Poisson distribution with $\lambda^{\mathrm{session}}_{\mathrm{e}}$ [calls/s], $\lambda^{\mathrm{session}}_{\mathrm{gin}}$ [calls/s], and $\lambda^{\mathrm{session}}_{\mathrm{gout}}$ [calls/s], respectively. Suppose that the mean call time follows an exponential distribution with $\frac{1}{\mu^{\mathrm{session}}_{\mathrm{e}}}$ [s], $\frac{1}{\mu^{\mathrm{session}}_{\mathrm{gin}}}$ [s], and $\frac{1}{\mu^{\mathrm{session}}_{\mathrm{gout}}}$ [s]. The traffic intensity is defined as $\rho_{\mathrm{e}}=\frac{\lambda^{\mathrm{session}}_{\mathrm{e}}}{N\mu^{\mathrm{session}}_{\mathrm{e}}}$, $\rho_{\mathrm{gin}}=\frac{\lambda^{\mathrm{session}}_{\mathrm{gin}}}{N\mu^{\mathrm{session}}_{\mathrm{gin}}}$, and $\rho_{\mathrm{gout}}=\frac{\lambda^{\mathrm{session}}_{\mathrm{gout}}}{N\mu^{\mathrm{session}}_{\mathrm{gout}}}$ using these values.

\subsection{M/M/s/s model}

\subsubsection{State transition diagram}
\begin{figure}[htbp]
    \centering
    \includegraphics[clip,width=12cm,height=65.0mm]{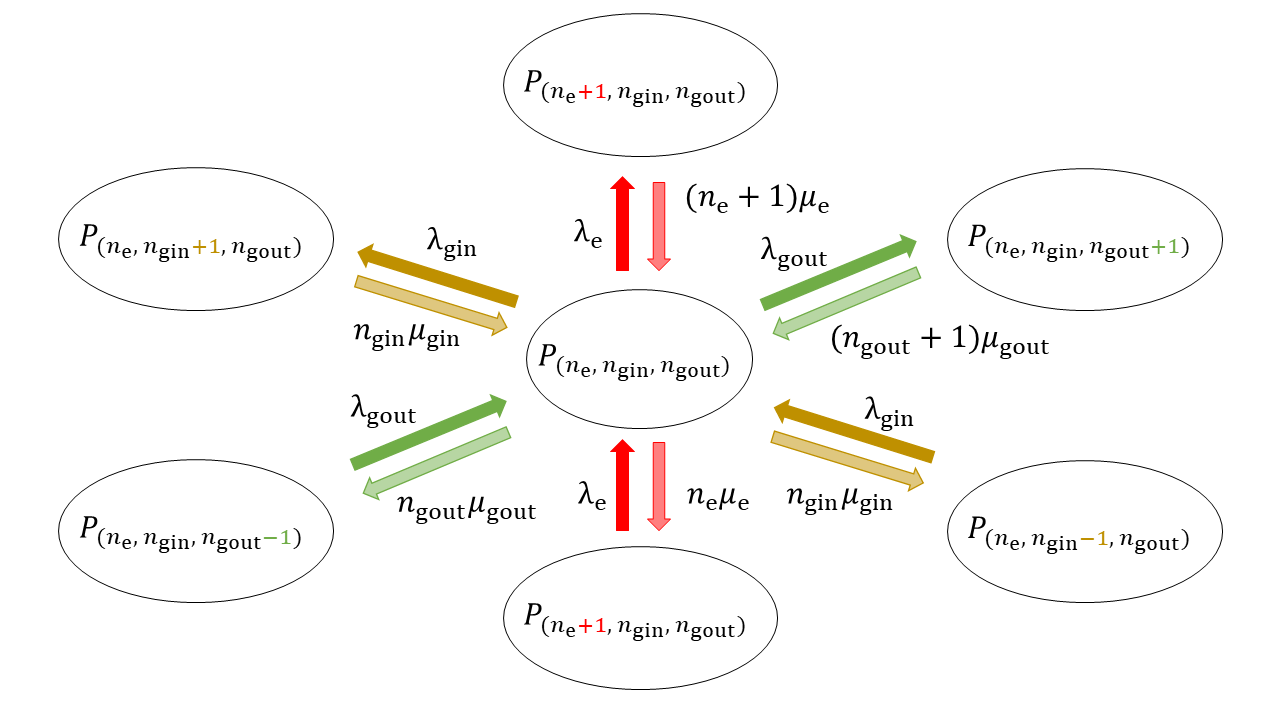}
    \caption{State transition diagram for situation A ($N_{\mathrm{now}} < t_{\mathrm{gout}}$).}
    \label{stateA}
\end{figure}

\begin{figure}[htbp]
    \centering
    \includegraphics[clip,width=12cm,height=65.0mm]{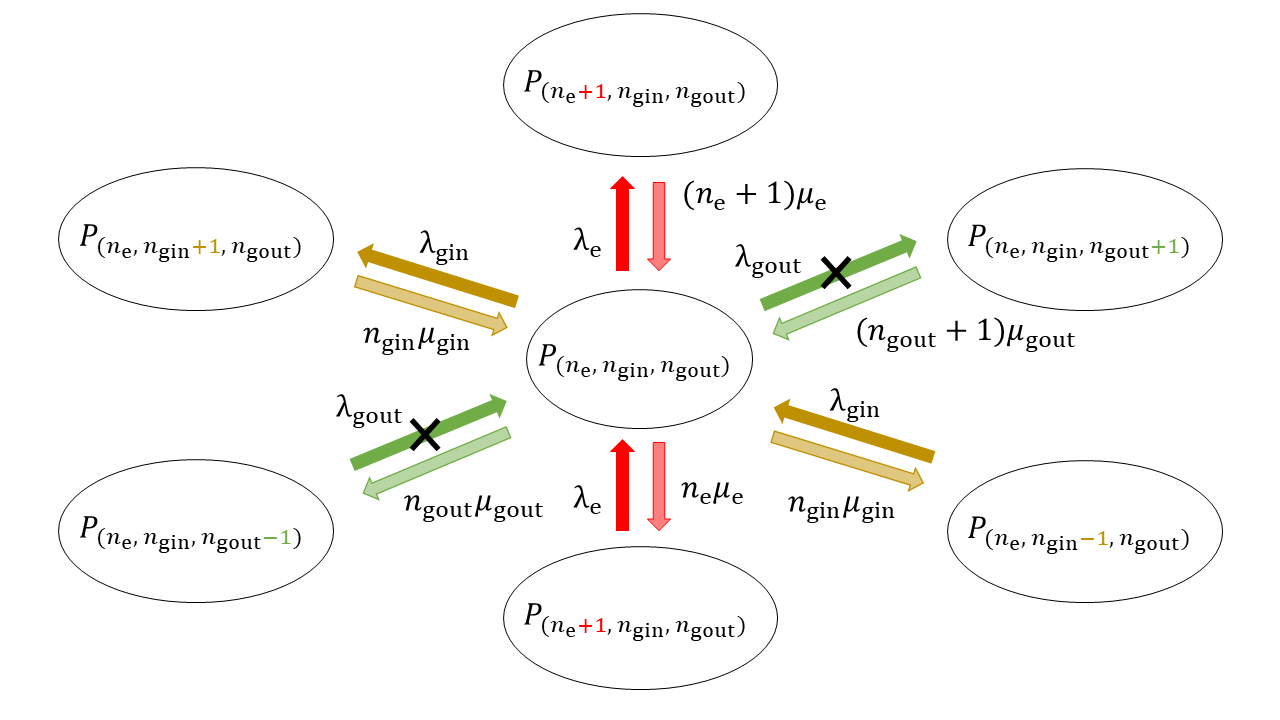}
    \caption{State transition diagram for situation B ($t_{\mathrm{gout}} \leq N_{\mathrm{now}} < t_{\mathrm{gin}}$).}
    \label{stateB}
\end{figure}

\begin{figure}[htbp]
    \centering
    \includegraphics[clip,width=12cm,height=65.0mm]{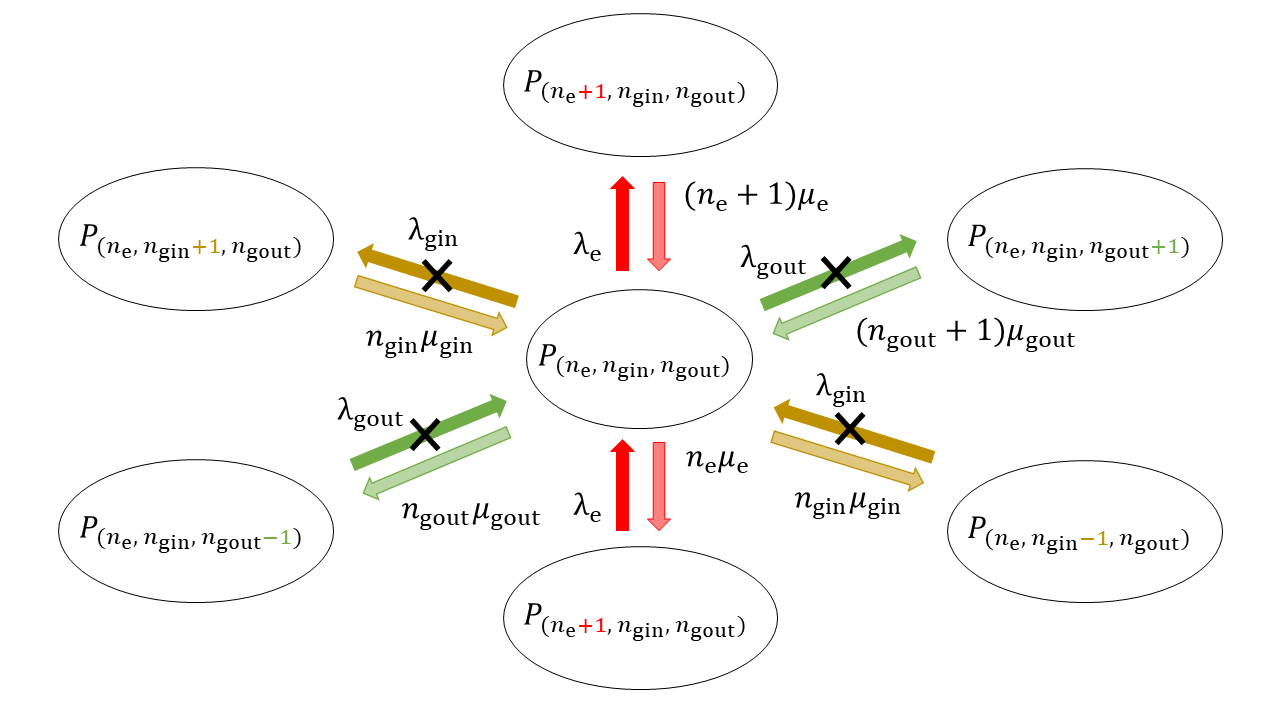}
    \caption{State transition diagram for situation C ($t_{\mathrm{gin}} \leq N_{\mathrm{now}}$).}
    \label{stateC}
\end{figure}

The arrival or departure transition of each VoIP session is shown in Figs. \ref{stateA}--\ref{stateC}. The state transition diagram of the call level can be classified into three situations using a threshold \cite{Kawase}. There are three situations: when the threshold is not needed when the threshold $t_{\mathrm{gout}}$ is needed to block the general-out VoIP session, and when the two thresholds $(t_{\mathrm{gin}}, t_{\mathrm{gout}})$ are needed to control the general-in VoIP session and the general-out VoIP session. In addition, we assume that each state has a steady state.

\begin{itemize}
    \item (Situation A)

        Situation A ($N_{\mathrm{now}} < t_{\mathrm{gout}}$) does not need the thresholds. Therefore, transition probabilities such as the arrival and departure are not restricted.
        
    \item (Situation B)
    
        Situation B ($t_{\mathrm{gout}} \leq N_{\mathrm{now}} < t_{\mathrm{gin}}$) restricts the general-out VoIP sessions using $t_{\mathrm{gout}}$. Therefore, the arrival transition of general-out VoIP sessions does not occur.
        
    \item (Situation C)
    
        Situation C ($t_{\mathrm{gin}} \leq N_{\mathrm{now}}$) restricts general-in VoIP sessions and general-out VoIP sessions using $t_{\mathrm{gin}}$ and $t_{\mathrm{gout}}$. Therefore, the arrival transition of general-out and general-in VoIP sessions does not occur.
        
\end{itemize}

\subsubsection{State equation}
The state equation of the call level can be derived using the state transition diagram in Figs. \ref{stateA}--\ref{stateC} \cite{Kawase}.

\begin{itemize}
    \item State A ($N_{\mathrm{now}} < t_{\mathrm{gout}}$)
        \begin{equation}
            \begin{split}  &(\lambda^{\mathrm{session}}_{\mathrm{e}}+\lambda^{\mathrm{session}}_{\mathrm{gin}}+\lambda^{\mathrm{session}}_{\mathrm{gout}}+n_{\mathrm{e}}\mu^{\mathrm{session}}_{\mathrm{e}}+n_{\mathrm{gin}}\mu^{\mathrm{session}}_{\mathrm{gin}} \quad+n_{\mathrm{gout}}\mu^{\mathrm{session}}_{\mathrm{gout}})P^{\mathrm{session}}_{(n_{\mathrm{e}},n_{\mathrm{gin}},n_{\mathrm{gout}})}\\
            &=\mu^{\mathrm{session}}_{\mathrm{e}}(n_{\mathrm{e}}+1)P^{\mathrm{session}}_{(n_{\mathrm{e}}+1,n_{\mathrm{gin}},n_{\mathrm{gout}})}+\mu^{\mathrm{session}}_{\mathrm{gin}}(n_{\mathrm{gin}}+1)P^{\mathrm{session}}_{(n_{\mathrm{e}},n_{\mathrm{gin}}+1,n_{\mathrm{gout}})}
            \\&\ \ \ +\mu^{\mathrm{session}}_{\mathrm{gout}}(n_{\mathrm{gout}}+1)P^{\mathrm{session}}_{(n_{\mathrm{e}},n_{\mathrm{gin}},n_{\mathrm{gout}}+1)}+\lambda^{\mathrm{session}}_{\mathrm{e}}P^{\mathrm{session}}_{(n_{\mathrm{e}}-1,n_{\mathrm{gin}},n_{\mathrm{gout}})}
            \\&\ \ \ \ \ \ +\lambda^{\mathrm{session}}_{\mathrm{gin}}P^{\mathrm{session}}_{(n_{\mathrm{e}},n_{\mathrm{gin}}-1,n_{\mathrm{gout}})}+\lambda^{\mathrm{session}}_{\mathrm{gout}}P^{\mathrm{session}}_{(n_{\mathrm{e}},n_{\mathrm{gin}},n_{\mathrm{gout}}-1)}.
            \end{split}
        \end{equation}

    \item Situation B ($t_{\mathrm{gout}} \leqq N_{\mathrm{now}} < t_{\mathrm{gin}}$) 

        \begin{equation}
            \begin{split}    &(\lambda^{\mathrm{session}}_{\mathrm{e}}+\lambda^{\mathrm{session}}_{\mathrm{gin}}+n_{\mathrm{e}}\mu^{\mathrm{session}}_{\mathrm{e}}+n_{\mathrm{gin}}\mu^{\mathrm{session}}_{\mathrm{gin}}+n_{\mathrm{gout}}\mu^{\mathrm{session}}_{\mathrm{gout}})P^{\mathrm{session}}_{(n_{\mathrm{e}},n_{\mathrm{gin}},n_{\mathrm{gout}})} \\
            &=\mu^{\mathrm{session}}_{\mathrm{e}}(n_{\mathrm{e}}+1)P^{\mathrm{session}}_{(n_{\mathrm{e}}+1,n_{\mathrm{gin}},n_{\mathrm{gout}})}+\mu^{\mathrm{session}}_{\mathrm{gin}}(n_{\mathrm{gin}}+1)P^{\mathrm{session}}_{(n_{\mathrm{e}},n_{\mathrm{gin}}+1,n_{\mathrm{gout}})}\\
            &\quad+\mu^{\mathrm{session}}_{\mathrm{gout}}(n_{\mathrm{gout}}+1)P^{\mathrm{session}}_{(n_{\mathrm{e}},n_{\mathrm{gin}},n_{\mathrm{gout}}+1)}+\lambda^{\mathrm{session}}_{\mathrm{e}}P^{\mathrm{session}}_{(n_{\mathrm{e}}-1,n_{\mathrm{gin}},n_{\mathrm{gout}})}+\lambda^{\mathrm{session}}_{\mathrm{gin}}P^{\mathrm{session}}_{(n_{\mathrm{e}},n_{\mathrm{gin}}-1,n_{\mathrm{gout}})}.
            \end{split}
        \end{equation}
    
    \item Situation C ($t_{\mathrm{gin}} \leqq N_{\mathrm{now}}$)

        \begin{equation}
            \begin{split}    &(\lambda^{\mathrm{session}}_{\mathrm{e}}+n_{\mathrm{e}}\mu^{\mathrm{session}}_{\mathrm{e}}+n_{\mathrm{gin}}\mu^{\mathrm{session}}_{\mathrm{gin}}+n_{\mathrm{gout}}\mu^{\mathrm{session}}_{\mathrm{gout}})P^{\mathrm{session}}_{(n_{\mathrm{e}},n_{\mathrm{gin}},n_{\mathrm{gout}})} \\
            &=\mu^{\mathrm{session}}_{\mathrm{e}}(n_{\mathrm{e}}+1)P^{\mathrm{session}}_{(n_{\mathrm{e}}+1,n_{\mathrm{gin}},n_{\mathrm{gout}})}+\mu^{\mathrm{session}}_{\mathrm{gin}}(n_{\mathrm{gin}}+1)P^{\mathrm{session}}_{(n_{\mathrm{e}},n_{\mathrm{gin}}+1,n_{\mathrm{gout}})}\\
            &\quad+\mu^{\mathrm{session}}_{\mathrm{gout}}(n_{\mathrm{gout}}+1)P^{\mathrm{session}}_{(n_{\mathrm{e}},n_{\mathrm{gin}},n_{\mathrm{gout}}+1)}+\lambda^{\mathrm{session}}_{\mathrm{e}}P^{\mathrm{session}}_{(n_{\mathrm{e}}-1,n_{\mathrm{gin}},n_{\mathrm{gout}})}.
            \end{split}
        \end{equation}
\end{itemize}

Let $\mathcal{D}$ be the set of all possible combinations of VoIP sessions that can exist in all situations A, B, and C. The following equation holds because the sum over all states is 1.\\
\begin{equation}
      \sum_{(n_{\mathrm{e}},n_{\mathrm{gin}},n_{\mathrm{gout}})\in \mathcal{D}}P^{\mathrm{session}}_{{\bm n}_a}= 1.
 \end{equation}

 \subsubsection{Call blocking probability}
The call blocking probability of VoIP sessions is derived by the steady-state probability of the call level $P^{\mathrm{session}}_{{\bm n}_a}$ of a VoIP session in the VoIP network. Arriving emergency VoIP sessions are blocked when $N_{\mathrm{now}}$ reaches $N$ or when the packet dropping probability exceeds the upper bound. Arriving general-out VoIP sessions are blocked when $t_{\mathrm{gout}} < N_{\mathrm{now}}$. Arriving general-in VoIP sessions are blocked when $t_{\mathrm{gin}} < N_{\mathrm{now}}$. The call blocking probability of each VoIP session is derived with the following equation.
  
\begin{equation}
L^{\mathrm{b}}_{\mathrm{e}}=\sum^{t_{\mathrm{gin}}}_{n_{\mathrm{gin}}=0}\sum^{t_{\mathrm{gout}}}_{n_{\mathrm{gout}}=0}P^{\mathrm{session}}_{(N-n_{\mathrm{gin}}-n_{\mathrm{gout}},n_{\mathrm{gin}}, n_{\mathrm{gout}})},
\end{equation}

\begin{equation}
 L^{\mathrm{b}}_{\mathrm{gin}}=L^{\mathrm{b}}_{\mathrm{e}}+\sum^{t_{\mathrm{gin}}}_{n_{\mathrm{gin}}=0}\sum^{t_{\mathrm{gout}}}_{n_{\mathrm{gout}}=0}P^{\mathrm{session}}_{(t_{\mathrm{gin}}-n_{\mathrm{gin}}-n_{\mathrm{gout}},n_{\mathrm{gin}},n_{\mathrm{gout}})},
\end{equation}

\begin{equation}
L^{\mathrm{b}}_{\mathrm{gout}}=L^{\mathrm{b}}_{\mathrm{gin}}+\sum^{t_{\mathrm{gin}}}_{n_{\mathrm{gin}}=0}\sum^{t_{\mathrm{gout}}}_{n_{\mathrm{gout}}=0}P^{\mathrm{session}}_{(t_{\mathrm{gout}}-n_{\mathrm{gin}}-n_{\mathrm{gout}},n_{\mathrm{gin}},n_{\mathrm{gout}})}.
\end{equation}
 
The call blocking probability of each VoIP session is obtained by the sum of $P^{\mathrm{session}}_{{\bm n}_a}$ when each VoIP session cannot be accommodated. 
However, because these state equations are difficult to solve \cite{miyata}, $P^{\mathrm{session}}_{{\bm n}_a}$ is calculated using numerical calculation for numerical analysis.

\subsection{MMPP/M/1/K model}

\subsubsection{MMPP}
\begin{figure}[htb]
    \centering
    \includegraphics[clip,width=11cm,height=60.0mm]{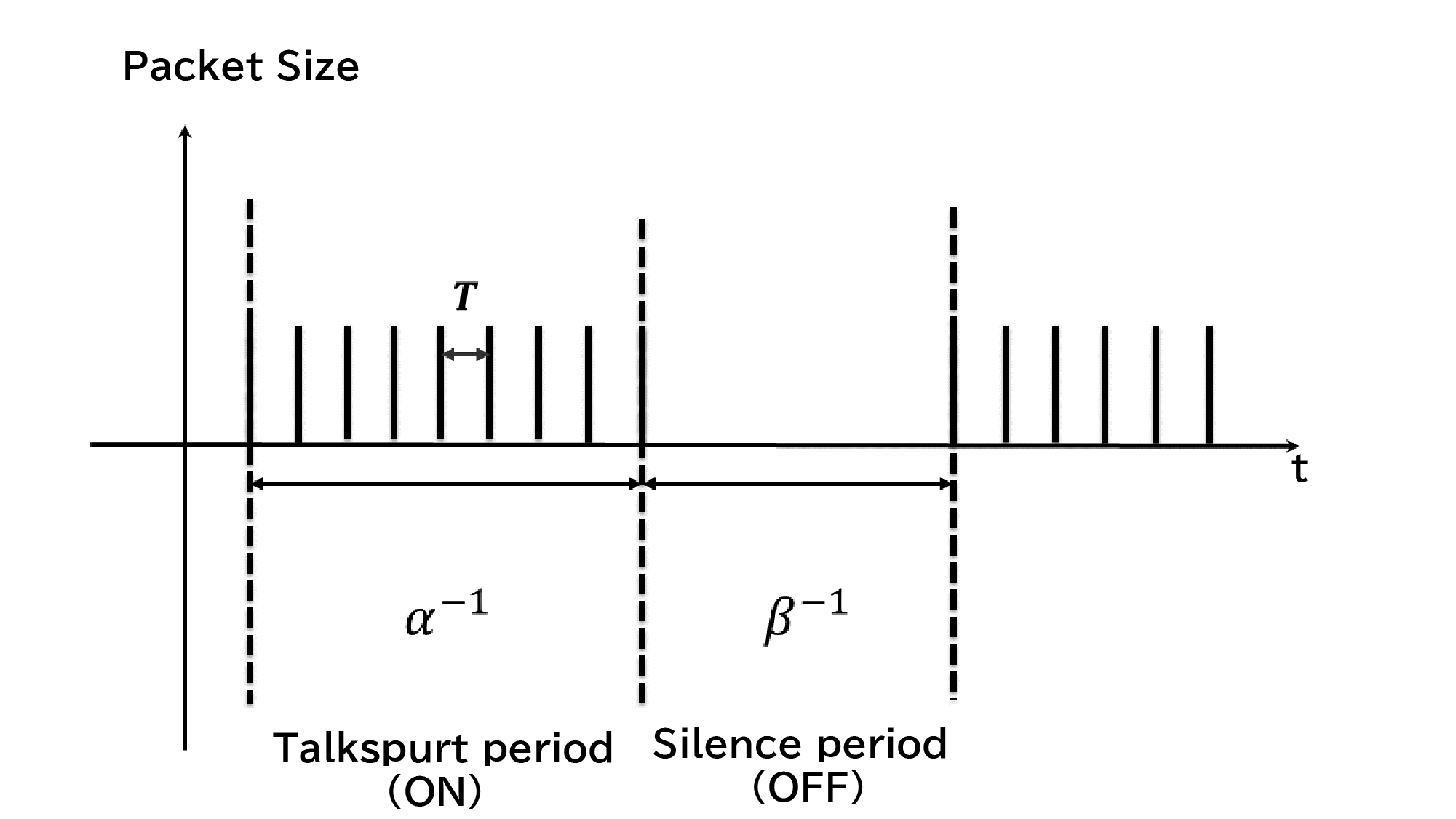}
    \caption{Arrival pattern of voice packets for single VoIP session \cite{Narikiyo}.}
    \label{arrival_packet}
\end{figure}

\begin{figure}[htb]
    \centering
    \includegraphics[clip,width=11cm,height=60.0mm]{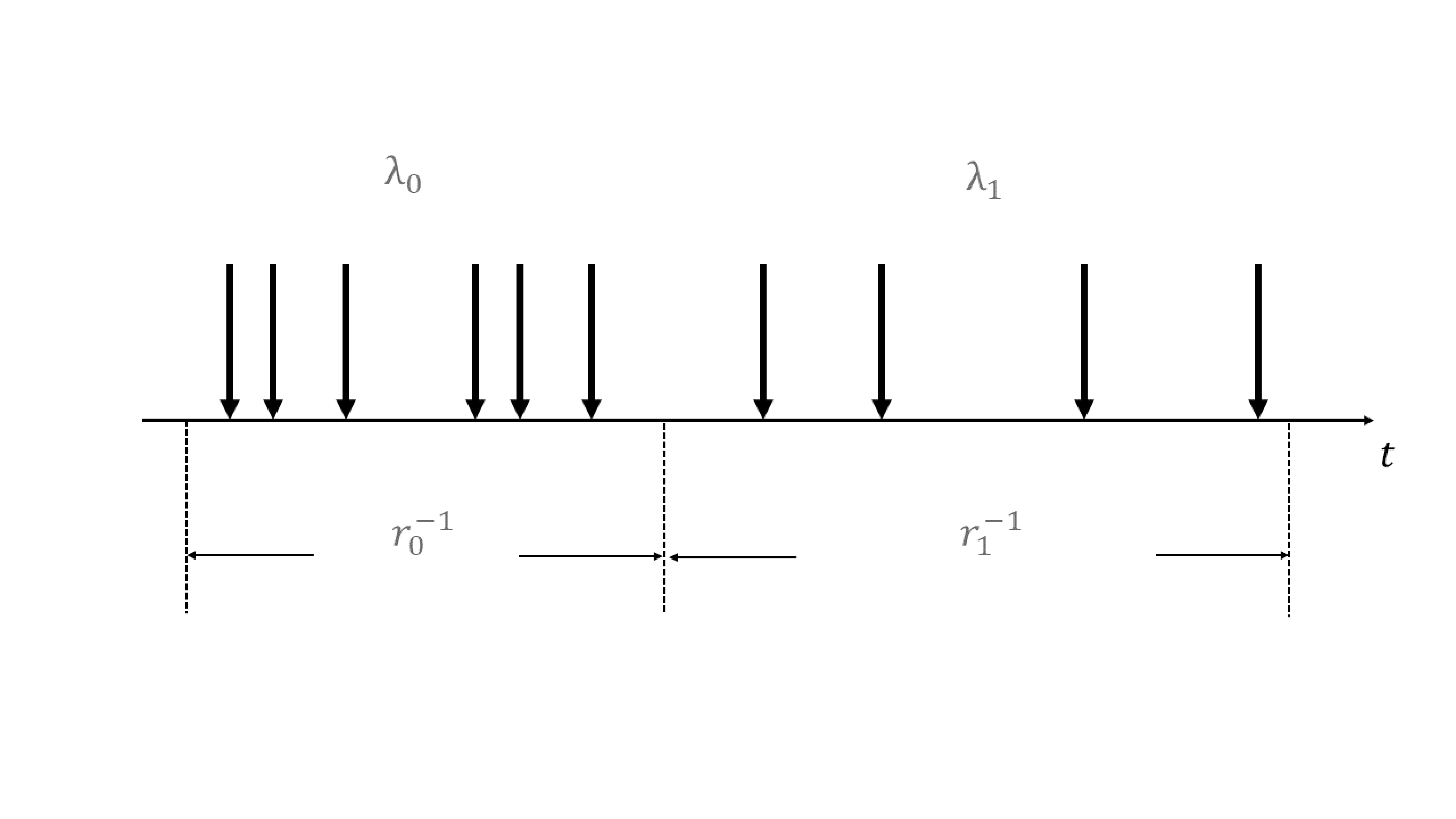}
    \caption{2-phase MMPP model.}
    \label{MMPP}
\end{figure}

Fig. \ref{arrival_packet} shows the arrival of voice packets, which are classified into talkspurt and silent period the basis of  a single VoIP session. The duration of these periods is assumed to follow an exponential distribution with means of $\alpha^{-1}$ and $\beta^{-1}$ for talkspurt and silence periods, respectively. We assume that the packet size $z$ and arrival interval $T$ are constant, and that packets arrive at intervals of $T$ during the talkspurt period and do not arrive at all during the silent period \cite{OnOff}. The packet arrival rate $\lambda_{\mathrm{p}}$, coefficient of square variation (variance/mean) $C^{2}{\mathrm{a}}$, and skewness (3rd order central product ratio/$\mathrm{variance}^{3/2}$) $S{\rm k}$ for this process are expressed as follows \cite{Akimaru}:

\begin{equation}
\lambda_{\mathrm{p}}=\frac{\beta}{T(\alpha+\beta)},
\label{packet_arrive}
\end{equation}

\begin{equation}
C^{2}_{\mathrm{a}}=\frac{1-(1-{\alpha}T)^{2}}{T^{2}(\alpha+\beta)^{2}},
\label{c2}
\end{equation}

\begin{equation}
S_{\mathrm{k}}=\frac{2{\alpha}T({\alpha}^{2}T^{2}-3{\alpha}T+3)}{[{\alpha}T(2-{\alpha}T)]^{3/2}}.
\label{Sk}
\end{equation}
Let $n$ be the number of VoIP sessions for one class. Fig. \ref{MMPP} shows the superposition process, in which packet arrivals from one class of VoIP session are superimposed for $n$ VoIP sessions. The packet mean inter-arrival time for the dense phase is $r^{-1}_{0}$ and the packet mean inter-arrival time for the sparse phase is $r^{-1}_{1}$. The subscripts 0 and 1 represent the dense and sparse phases. The mean duration of each phase follows an exponential distribution. There are two phases with packet arrival rates $\lambda^{\mathrm{packet}}_{0}$ and $\lambda^{\mathrm{packet}}_{1}$, which is a Poisson process \cite{Matsuoka}. This is called two-phase MMPP. Let $q_{0}$ be the transition probabilities of the phase, which are from dense phase to sparse phase, and  $q_{1}$ be the transition probabilities of the phase, which are from sparse phase to dense phase. Each transition probability of the phases $q_{0}$, $q_{1}$ and each packet arrival rate of the phases $\lambda^{\mathrm{packet}}_{0}$, $\lambda^{\mathrm{packet}}_{1}$ are represented by the following equations using $\lambda_{\mathrm{p}}$, $C^{2}_{\mathrm{a}}$, $S_{\mathrm{k}}$, and $n$.

\begin{equation}
\left.
\begin{array}{l}
q_{0} \\
q_{1}
\end{array}
\right\} \\
=D(1\pm 1/\sqrt{1+n\lambda_{\mathrm{p}} E)},
\label{r01}
\end{equation}

\begin{equation}
\left.
\begin{array}{l}
\lambda^{\mathrm{packet}}_{0} \\
\lambda^{\mathrm{packet}}_{1}
\end{array}
\right\} \\
=n\lambda_{\mathrm{p}}+F\pm F\sqrt{1+n\lambda_{\mathrm{p}} E},
\label{lambdas}
\end{equation}
where
\begin{equation}
    D=\frac{3\lambda_{\mathrm{p}}(C^2_{\mathrm{a}}-1)}{2S_{\mathrm{k}}C^{3}_{\mathrm{a}}-3C^{4}_{\mathrm{a}}-1},
\label{D}
\end{equation}
\begin{equation}
    E=D\frac{C^2_{\mathrm{a}}-1}{F^2},
\label{E}
\end{equation}
\begin{equation}
    F=D\frac{3C^{4}_{\mathrm{a}}-S_{\mathrm{k}}C^{3}_{\mathrm{a}}-3C^{2}_{\mathrm{a}}+2}{3(C^{2}_{\mathrm{a}}-1)}.
\label{F}
\end{equation}

Let $Q_{\mathrm{single}}$ be an infinitesimal generating operator of a continuous-time Markov chain for one class of VoIP session, and the diagonal elements of $\Lambda^{\mathrm{packet}}_{\mathrm{single}}$ be the packet arrival rate at each phase of the one class of VoIP session.
Using $q_{0}$, $q_{1}$, $\lambda^{\mathrm{packet}}_{0}$, and $\lambda^{\mathrm{packet}}_{1}$, the two-phase MMPP is represented by the following two square matrices $Q_{\mathrm{single}}$ and $\Lambda^{\mathrm{packet}}_{\mathrm{single}}$.

\begin{equation}
    Q_{\mathrm{single}}=
    \begin{pmatrix}
    -q_0 & q_0 \\
    q_1 & -q_1 \\
    \end{pmatrix},
\label{single_Q}
\end{equation}

\begin{equation}
    \Lambda^{\mathrm{packet}}_{\mathrm{single}}=
    \begin{pmatrix}
    \lambda^{\mathrm{packet}}_{0} & 0 \\
    0 & \lambda^{\mathrm{packet}}_{1} \\
    \end{pmatrix}.
\label{single_lambda}
\end{equation}

Moreover, the superposition process of multiple MMPPs is also MMPP \cite{Kasahara}. Let $Q_i$, $\Lambda_i^{\mathrm{packet}}$ be $Q_{\mathrm{single}}$, $\Lambda_{\mathrm{single}}^{\mathrm{packet}}$ for VoIP session type $i$ $(1\leq i \leq c)$. The parameter $\bm{M}_i=$ ($Q_{i}$, $\Lambda^{\mathrm{packet}}_{i}$) ($1\leq i \leq c$) represents $c$ kinds of MMPP. Thus, $\bm{M}=$ ($Q$, $\Lambda^{\mathrm{packet}}$) is given by the following formula, and the number of phase states is $2^c$.

\begin{equation}
Q=Q_{\mathrm{1}}\oplus Q_{\mathrm{2}}\oplus ...\oplus Q_{c},
\label{several_Q}
\end{equation}
\begin{equation}
\Lambda^{\mathrm{packet}}=\Lambda^{\mathrm{packet}}_{\mathrm{1}}\oplus \Lambda^{\mathrm{packet}}_{\mathrm{2}}\oplus ...\oplus \Lambda^{\mathrm{packet}}_{c}.
\label{several_lambda}
\end{equation}
where $\oplus$ represents the Kronecker sum in a square matrix.

 In this study, each VoIP session is classified into three classes: emergency VoIP sessions, general-in VoIP sessions, and general-out VoIP sessions. Therefore, three 2-phase MMPPs are superimposed, resulting in a $2^3$=8 phase MMPP. Therefore, the number of accommodated VoIP sessions on $n$ in Eqs. (\ref{r01}) and (\ref{lambdas}) is $n_{\mathrm{e}}, n_{\mathrm{gin}}$, and $n_{\mathrm{gout}}$ for each of the respective classes. 
 From the above, using Eqs. (\ref{several_Q}) and (\ref{several_lambda}), we obtain 

\begin{equation}
Q=Q_{\mathrm{e}}\oplus Q_{\mathrm{gin}}\oplus Q_{\mathrm{gout}},
\label{8isou_Q}
\end{equation}
\begin{equation}
\Lambda^{\mathrm{packet}}=\Lambda^{\mathrm{packet}}_{\mathrm{e}}\oplus \Lambda^{\mathrm{packet}}_{\mathrm{gin}}\oplus \Lambda^{\mathrm{packet}}_{\mathrm{gout}}.
\label{8isou_lambda}
\end{equation}

\subsubsection{State transition diagram}
\begin{figure}[htbp]
  \begin{center}
  \hspace{-6mm}
    \includegraphics[clip,width=12cm,height=65.0mm]{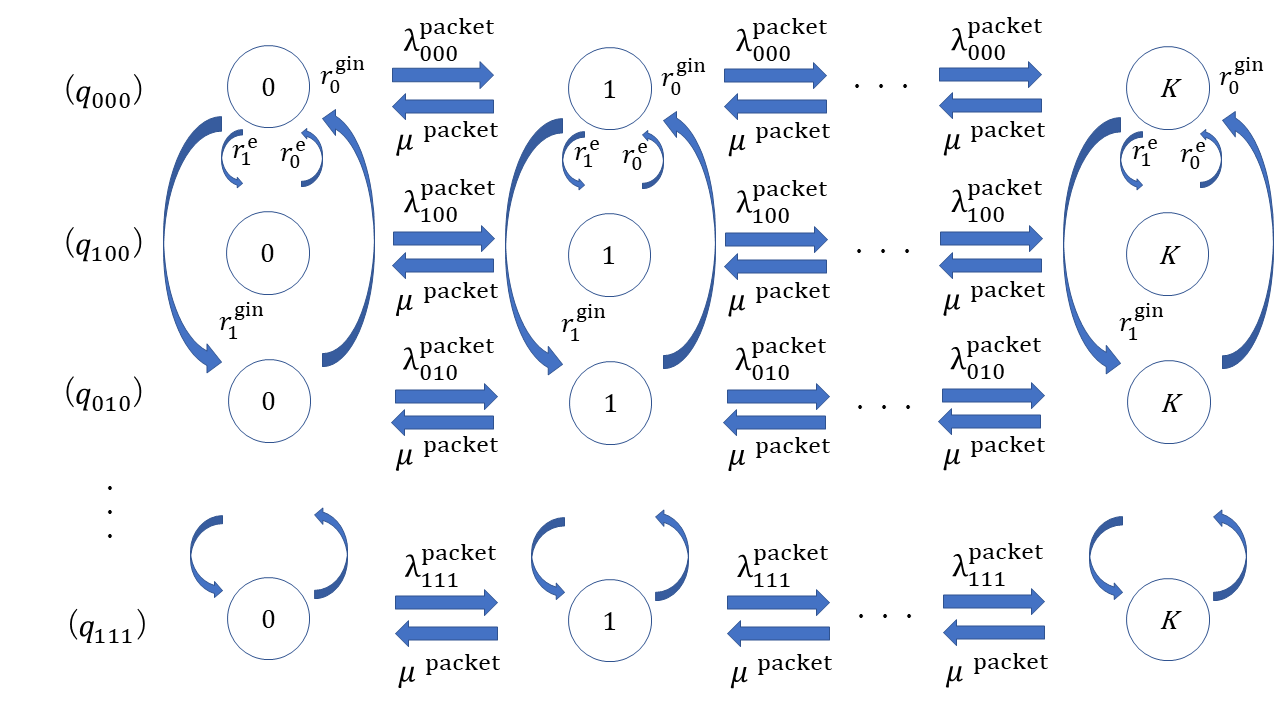}
    \caption{State transition diagram of packet level for three classes of VoIP sessions.}
    \label{fig_packet}
  \end{center}
\end{figure}

Fig. \ref{fig_packet} shows the state transition diagram of the packet level.  Let $\mu^{\mathrm{packet}}$ be the mean packet processing rate. We assume that voice packets consisting of multiple VoIP sessions continuously arrive at a single control server whose average processing time  $\frac{1}{\mu^{\mathrm{packet}}}$ follows an exponential distribution. For this purpose, we model the situation using the $MMPP/M/1/K$ model of queueing theory. The state transition diagram in Fig. \ref{fig_packet}  represents the dense and sparse phases of three classes of VoIP session by dividing the packet into eight phases. 

\subsubsection{Equation of state for eight-phase MMPP/M/1/K}
The packet arrival and processing for the three classes of VoIP sessions are represented as eight phases in the model  \cite{Narikiyo}. Let $q_{000}$, $q_{001}$, $q_{010}$, $q_{011}$, $q_{100}$, $q_{101}$, $q_{110}$, and $q_{111}$ be the eight-phase with the dense phase as 0 and the sparse phase as 1, and $\lambda^{\mathrm{packet}}_{ijk}$ be the arrival rate for the phase $q_{ijk} (i,j,k \in [0,1])$. In addition, let $K$ be the capacity of the control server's queue, and $P^{\mathrm{packet}}_{ijk}(m) (0\leq m\leq K)$ be the steady-state probability of the packet level when the number of packets waiting in the queue is $m$. The state equation of the packet level is:\\

\begin{align}
&P^{\rm packet}_{ijk}(m)(\lambda^{\mathrm{packet}}_{ijk}a_{K-m}+\mu^{\mathrm{packet}} a_{m}+q_{i}+q_{j}+q_{k})\notag
\\&=P^{\mathrm{packet}}_{ijk}(m-1)\lambda^{\mathrm{packet}}_{ijk}a_{m}+P^{\mathrm{packet}}_{ijk}(m+1)\mu a_{K-m}\notag
\\&\ \ \ \  +P^{\mathrm{packet}}_{ijk}(m)q_{i}+P^{\mathrm{packet}}_{ijk}(m)q_{j}+P^{\mathrm{packet}}_{ijk}(m)q_{k}\ \ (i,j,k\in [0,1]),
\label{packet_equation}
\end{align}
where $a_m$ is a variable that means that the control server does not process packets (does not let them arrive) when there are no packets in the queue (full) $m=0,(K)$. 

The sum of the steady-state probability is 1. Let $\mathcal M$ be the state space of the combination vector $(i,j,k,m)$ representing the states of the eight-phase, including the number of packets waiting  $m$ in the queue; then, the following equation holds:

 \begin{equation}
      \sum_{(i,j,k,m)\in \mathcal M}P^{\mathrm{packet}}_{ijk}(m) = 1.
\label{packet_state}
 \end{equation}

\subsubsection{Packet dropping probability}
A packet is dropped when a new packet arrives when the queue is full of packets. Let ${\mathcal S}$ be for the space of each phase $(i,j,k)$, and the packet dropping probability per VoIP session combination $L^{\mathrm{d}}({\bm n}_a)$ is derived by the following equation:

\begin{equation}
\begin{split}
L^{\mathrm{d}}({\bm n}_a) = \frac{\sum_{(i,j,k) \in \mathcal S} \lambda^{\mathrm{packet}}_{ijk}({\bm n}_a) P^{\mathrm{packet}}_{ijk}(K)}{\sum_{(i,j,k) \in \mathcal S} \lambda^{\mathrm{packet}}_{ijk}({\bm n}_a)}\quad({\bm n}_a \in \mathcal{N}).
\end{split}
\label{PacketLossProbability}
\end{equation}

 The packet dropping probability per VoIP session combination $L^{\mathrm{d}}({\bm n}_a)$ is present for each VoIP session combination (blue circle) in Fig. \ref{StateSpace_flow}.
 In this study, we set the packet dropping probability to an upper bound $\bar{L}^{\mathrm{d}}_{\mathrm{upper}}$. From Eq. (\ref{PacketLossProbability}), the average packet dropping probability $\Bar{L}^{\mathrm{d}}({\mathcal N}_{\bm t})$ for the total number of each VoIP session combination ${\mathcal N}_{\bm t}$ is derived by the following equation by using the steady-state probabilities of the call level $P^{\mathrm{session}}_{{\bm n}_a}$ and $L^{\mathrm{d}}({\bm n}_a)$.

    \begin{equation}
    \begin{split}
    \Bar{L}({\mathcal N}_{\bm t}) = \sum_{{\bm n}_a \in {\mathcal N}_{\bm t}} P^{\mathrm{session}}_{{\bm n}_a}L({\bm n}_a).
    \end{split} 
    \label{average_packetloss}
    \end{equation}
    
 The average packet dropping probability $\Bar{L}^{\mathrm{d}}({\mathcal N}_{\bm t})$ is the expected value for the overall state space of the call level in Figs. \ref{StateSpace_33} or \ref{StateSpace_21}. However, because these state equations are difficult to solve \cite{miyata}, $P^{\mathrm{session}}_{{\bm n}_a}$ and $L^{\mathrm{d}}({\bm n}_a)$ are calculated using numerical calculation for numerical analysis. Thus, we do a full search of $(t_{\mathrm{gin}}, t_{\mathrm{gout}})$ to obtain this $(t^*_{\mathrm{gin}}, t^*_{\mathrm{gout}})$  for each traffic condition. As a result, we can find $(t^*_{\mathrm{gin}}, t^*_{\mathrm{gout}})$ for each traffic condition.

 \section{Numerical analysis}
 \subsection{Parameter settings}
 
\begin{table}[H]
    \caption{Parameter Settings.}
    \centering
    \scalebox{0.75}{
    \begin{tabular}{c||c}
    \hline \hline
    Parameter & Input value \\
    \hline \hline
    Maximum number of VoIP sessions accommodated, $N$ & 20 \\ \hline
    Bandwidth, $B$ & 1.25 Mbps \\ \hline
    Mean packet size, $z$ & 1744 bit\\ \hline
    Talkspurt period, $\alpha^{-1}$ & 352 ms \\ \hline
    Silent period, $\beta^{-1}$ & 650 ms \\ \hline
    Departure rate of each VoIP session, ($\mu^{\mathrm{session}}_{\mathrm{e}}$, $\mu^{\mathrm{session}}_{\mathrm{gin}}$, $\mu^{\mathrm{session}}_{\mathrm{gout}}$) & 0.01 \\ \hline
    Upper bound call blocking probability for emergency VoIP sessions, $C_{\mathrm{e}}$ & 0.15 \\ \hline
    Upper bound average packet dropping probability, $\Bar{L}^{\mathrm d}_{\mathrm{upper}}$ & 0.0025 \\ \hline
    \hline
    \end{tabular}
    }
    \label{ParameterSet}
\end{table}

As shown in Table \ref{ParameterSet}, we set the parameters for numerical analysis as follows. Our numerical analysis in this study evaluated the packet dropping probability and the call blocking probability with a bandwidth of $B=$1.25 Mbps and a maximum number of accommodated VoIP sessions $N$=20. We assumes G.711 as the codec, thus meaning that the packet size is $z=1744$ bits \cite{Murakami}. The packet processing rate $\mu^{\mathrm{packet}}$ is derived by $B/z$ \cite{Lopez}. The value representing the arrival of each packet in one class of VoIP session is assumed to be the same value for emergency VoIP sessions, general-in VoIP sessions, and general-out VoIP sessions, and the talkspurt period $\alpha^{-1}=352$ ms, silent period $\beta^{-1}=650$ ms, and packet arrival interval $T=16$ ms \cite{Lopez}. In addition, to assume a mean talk time of 100 s, the departure rate for each VoIP session is $\mu^{\mathrm{session}}=0.01$ \cite{Kawase}. The traffic intensity of emergency VoIP sessions $\rho_{\mathrm{e}}$, general-in VoIP sessions $\rho_{\mathrm{gin}}$, and general-out VoIP sessions $\rho_{\mathrm{gout}}$ is assumed to be an emergency, so the total traffic intensity for each VoIP session is set to be around 2.0 \cite{Kawase}. The upper bound call blocking probability for emergency VoIP sessions $C_{\mathrm{e}}$ is defined by a specified value and is therefore 0.15 \cite{uppercallblock}. The upper bound call blocking probability for general-in VoIP sessions $C_{\mathrm{gin}}$ is set to 0.5.


In Sec. 5. 2), we show that the comparison of simulation and theoretical analysis. In this Sec. 5. 3), 4), 5), we analyze the CAC method when varying each traffic intensity $\rho_{\mathrm{e}}, \rho_{\mathrm{gin}}$, and $\rho_{\mathrm{gout}}$ for the three patterns. In Sec. 5. 3. 2), we perform a characteristic analysis of the packet dropping probability and call blocking probability to derive the optimal threshold. In addition, in Sec. 5. 4. 2), we perform an analysis by varying the upper bound call blocking probability for the general-in VoIP session $C_{\mathrm{gin}}$ to further analyze the key parameters of our proposed CAC method.

\subsection{Comparison of simulation and theoretical analysis}

\begin{figure}[ht]
    \centering
    \hspace{-7.5mm}
    \includegraphics[clip,width=12cm,height=90.0mm]{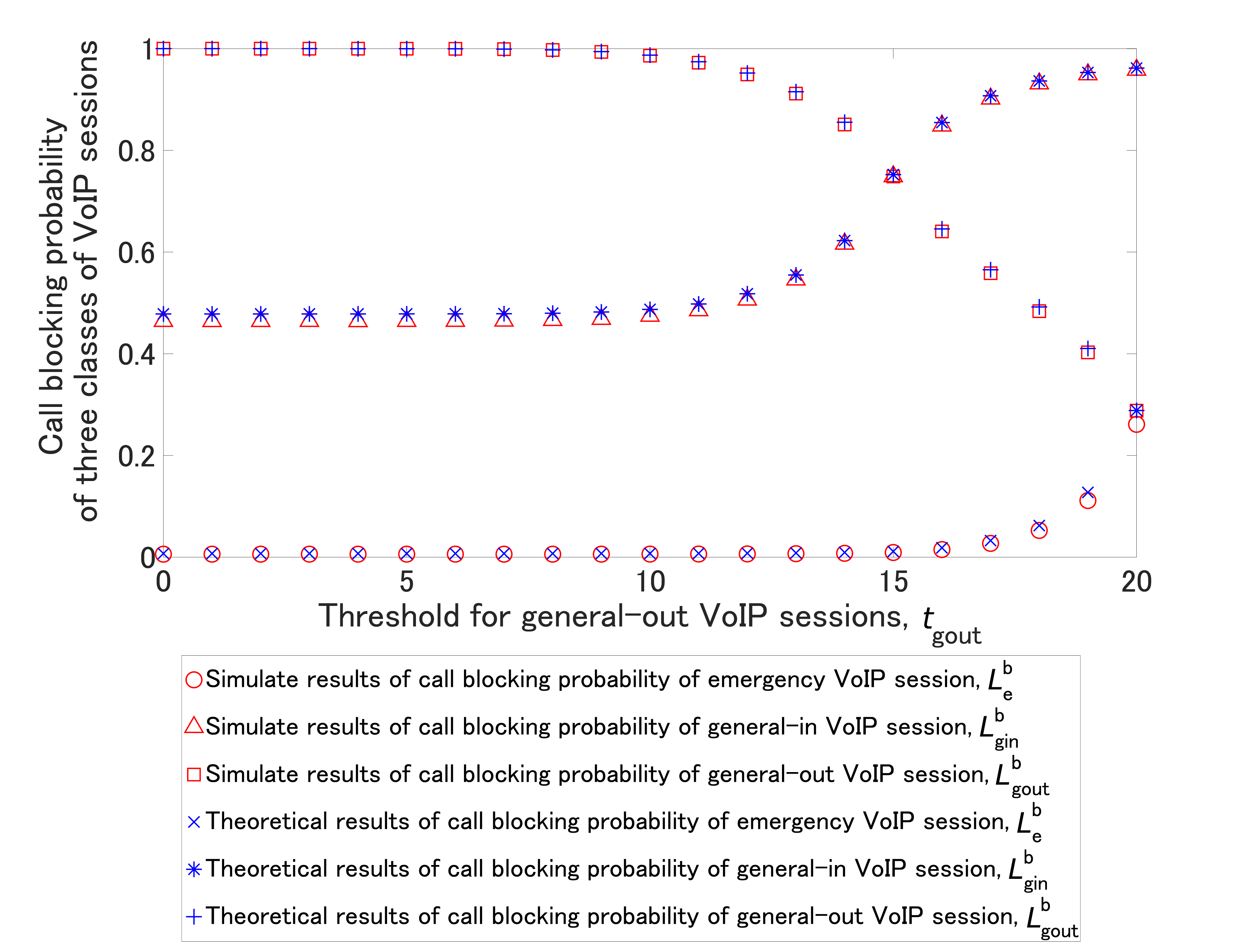}
    \caption{A comparison of simulation and theoretical results is made for the call blocking probability of three classes of VoIP sessions shown in Fig. \ref{allcallblock_thgout} in Sec. 5.3.2.}
    \label{SimuAndTheorerical}
\end{figure}

We perform simulations for the call level because the threshold varies significantly with the traffic load of call level. Therefore, we compare the simulation results with the theoretical results for the call blocking probability of three classes of VoIP sessions shown in Fig. \ref{allcallblock_thgout} in Sec. 5.3.2. The results are shown in Fig. \ref{SimuAndTheorerical}, so the red plots in Fig. \ref{SimuAndTheorerical} are the same results as in Sec. 5.3.2.

In the comparison of these results, the simulation results with the theoretical results show no significant differences. Therefore, the theoretical results of our proposed CAC method are shown to be valid.

\subsection{Analysis of CAC method when varying traffic intensity of emergency VoIP sessions}
\subsubsection{Comparison of conventional CAC method and our proposed CAC method when varying traffic intensity of emergency VoIP sessions}

\begin{figure}[ht]
    \centering
    \hspace{-7.5mm}
    \includegraphics[clip,width=12cm,height=65.0mm]{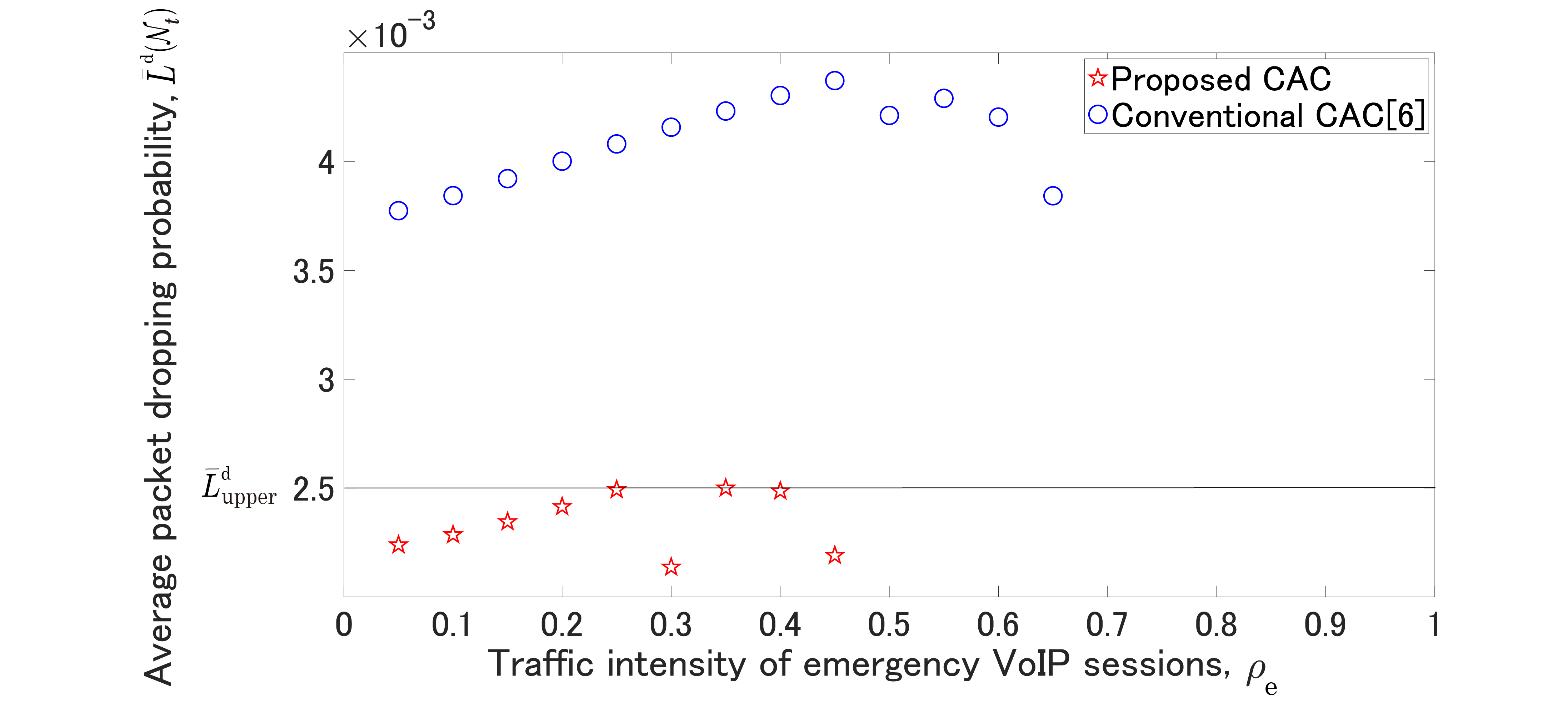}
    \caption{Traffic intensity of emergency VoIP sessions and average packet dropping probability.}
    \label{packetloss_e_12}
\end{figure}

\begin{figure}[ht]
    \centering
    \includegraphics[clip,width=11.5cm,height=65.0mm]{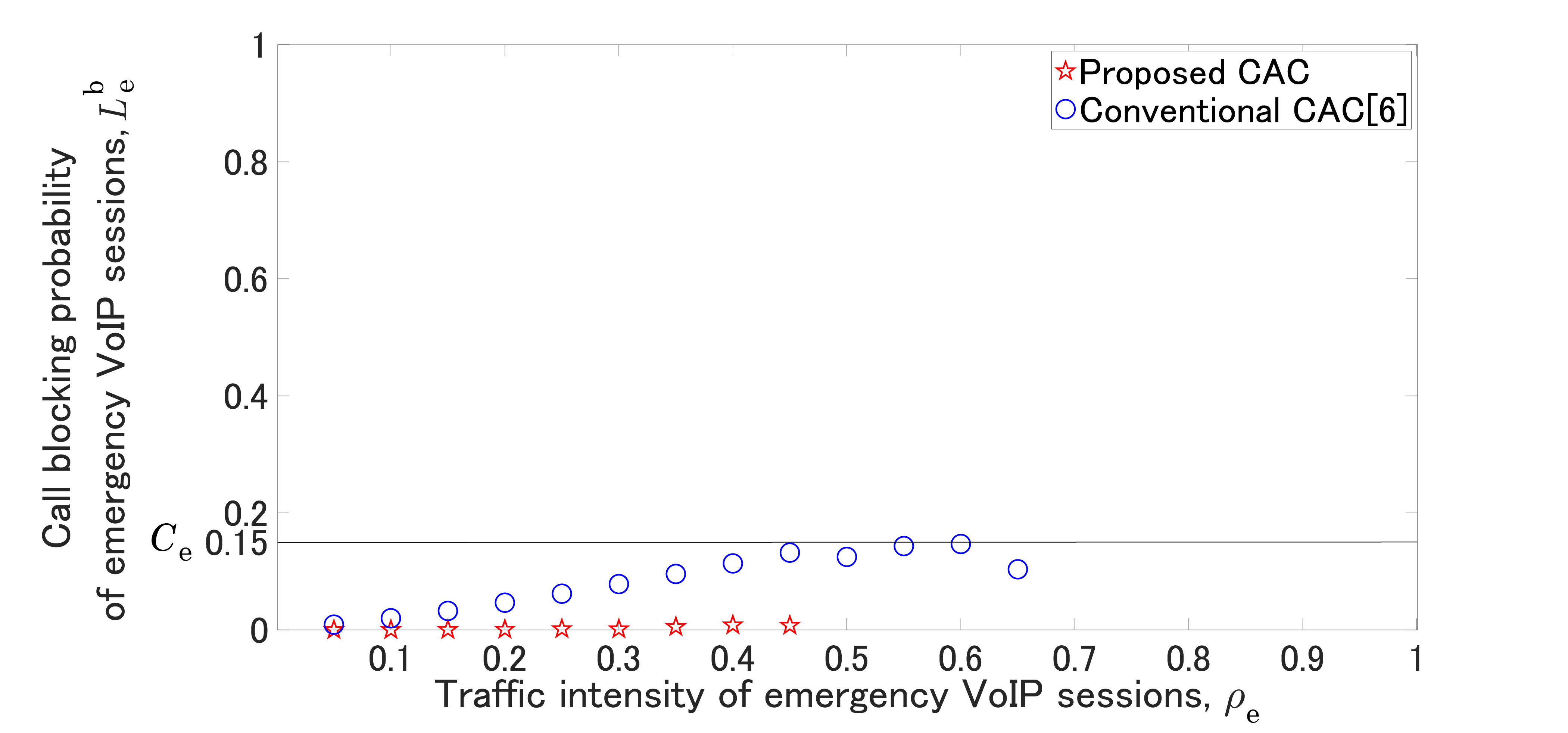}
    \caption{Traffic intensity of emergency VoIP sessions and call blocking probability of emergency VoIP session.}
    \label{emergency_block_e}
\end{figure}

\begin{figure}[ht]
    \centering
    \hspace{4.5mm}
    \includegraphics[clip,width=11cm,height=65.0mm]{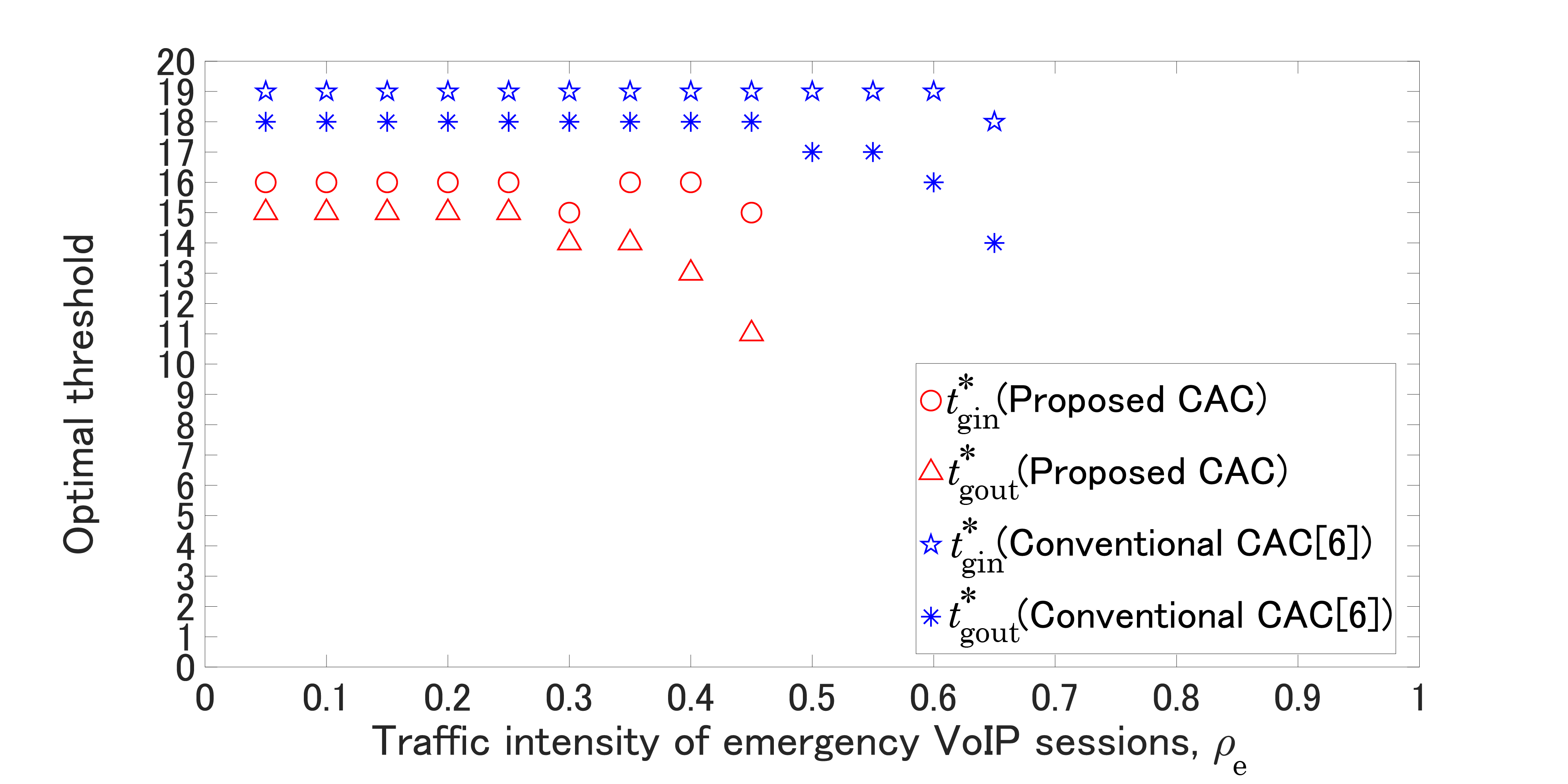}
    \caption{Traffic intensity of emergency VoIP sessions and each optimal threshold.}
    \label{threshold_e_12}
\end{figure}

Figs. \ref{packetloss_e_12}--\ref{threshold_e_12} shows a comparison of the evaluated values for the conventional method \cite{Kawase} and our proposed method when the traffic intensity of the emergency VoIP session $\rho_{\mathrm{e}}$ is varied from 0.05 to 1.0. The traffic intensity pattern is $(\rho_{\mathrm{e}},\rho_{\mathrm{gin}},\rho_{\mathrm{gout}})=(\rho_{\mathrm{e}},0.5,0.8)$. 

Fig. \ref{packetloss_e_12} shows the average packet dropping probability $\Bar{L}^{\mathrm{d}}({\mathcal N}_{\bm t})$ at the optimal thresholds $(t^*_{\mathrm{gin}}, t^*_{\mathrm{gout}})$. Fig. \ref{emergency_block_e} shows the call blocking probability of the emergency VoIP session $L^{\mathrm{b}}_{\mathrm{e}}$ at the optimal thresholds $(t^*_{\mathrm{gin}}, t^*_{\mathrm{gout}})$.  Fig. \ref{threshold_e_12} shows the optimal thresholds $(t^*_{\mathrm{gin}}, t^*_{\mathrm{gout}})$. As shown in Fig. \ref{packetloss_e_12}, while $\Bar{L}^{\mathrm{d}}({\mathcal N}_{\bm t})$ of the conventional method is 0.0035 or higher, our proposed method can reduce $\Bar{L}^{\mathrm{d}}({\mathcal N}_{\bm t})$ to 0.0025 or lower. Therefore, the optimal threshold ${\bm t}^*$ of our proposed method is able to guarantee the communication quality. As $\rho_{\mathrm{e}}$ increases, $\Bar{L}^{\mathrm{d}}({\mathcal N}_{\bm t})$ increases as well. This is because $\Bar{L}^{\mathrm{d}}({\mathcal N}_{\bm t})$ is considered by the steady-state probability of the call level $P^{\mathrm{session}}_{{\bm n}_a}$ in Eq. (\ref{average_packetloss}). Thus, $\Bar{L}^{\mathrm{d}}({\mathcal N}_{\bm t})$ depends on the traffic intensity of VoIP sessions. Note that traffic intensity patterns exist in which $\rho_{\mathrm{e}}$ increases where as $\Bar{L}^{\mathrm{d}}({\mathcal N}_{\bm t})$ decreases by $(t^*_{\mathrm{gin}}, t^*_{\mathrm{gout}})$ in Fig. \ref{threshold_e_12}.

As shown in Figs. \ref{packetloss_e_12} and \ref{threshold_e_12}, $\Bar{L}^{\mathrm{d}}({\mathcal N}_{\bm t})$ decreases at $\rho_{\mathrm{e}}$, where $(t^*_{\mathrm{gin}}, t^*_{\mathrm{gout}})$ becomes smaller. In general, as the number of accommodated VoIP sessions $N_{\mathrm{now}}$ increases, the packet dropping probability $L^{\mathrm{d}}({\bm n}_a)$ also increases \cite{Narikiyo}. The reason for the decrease in $\Bar{L}^{\mathrm{d}}({\mathcal N}_{\bm t})$ is that high $P^{\mathrm{session}}_{{\bm n}_a}$, such as high $N_{\mathrm{now}}$, does not exist when controlling at the optimal threshold ${\bm t}^*$. From this characteristic, the communication quality can be guaranteed by using the appropriate thresholds.

 As shown in Fig. \ref{emergency_block_e}, $L^{\mathrm{b}}_{\mathrm{e}}$ is almost zero for our proposed CAC method compared with the conventional CAC method \cite{Kawase}. This is because the characteristics of the steady-state probability are affected by the thresholds. Thus, we analyze the thresholds and characteristics of the call blocking probability and packet dropping probability in Sec. V. B. 2).

\subsubsection{Characteristic analysis of packet dropping probability and call blocking probability to derive optimal thresholds}

\begin{figure*}[ht]
    \begin{minipage}[t]{0.5\linewidth}
        \centering
        \includegraphics[clip,width=8cm,height=60mm]{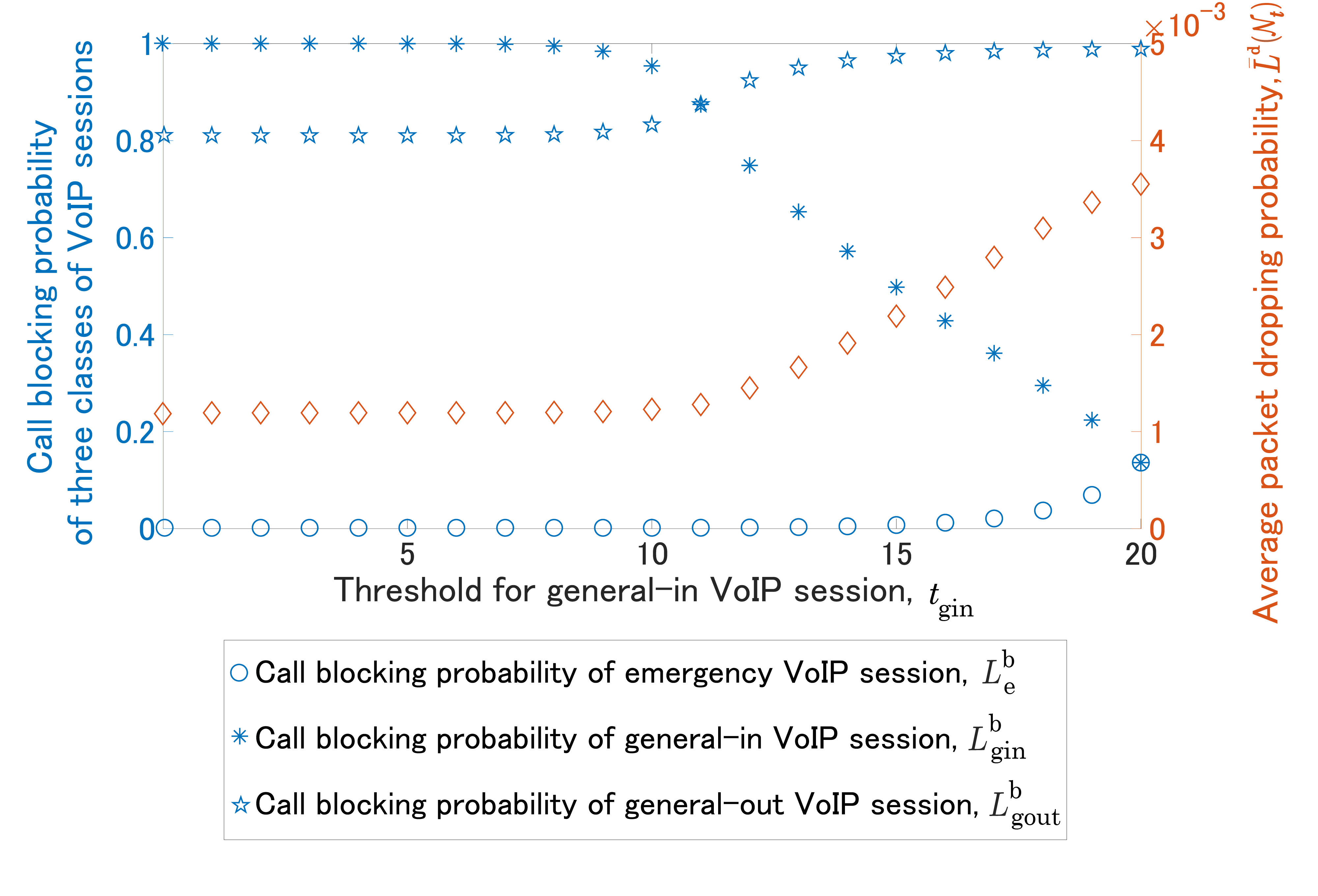}
        \subcaption{When optimal threshold for general-out VoIP session $t^*_{\mathrm{gout}}$ is 11, threshold for general-in VoIP session $t^*_{\mathrm{gin}}$ and each call blocking probability or average packet dropping probability.}
        \label{allcallblock_thgin}
    \end{minipage}
    \begin{minipage}[t]{0.5\linewidth}
        \centering
        \includegraphics[clip,width=8cm,height=60mm]{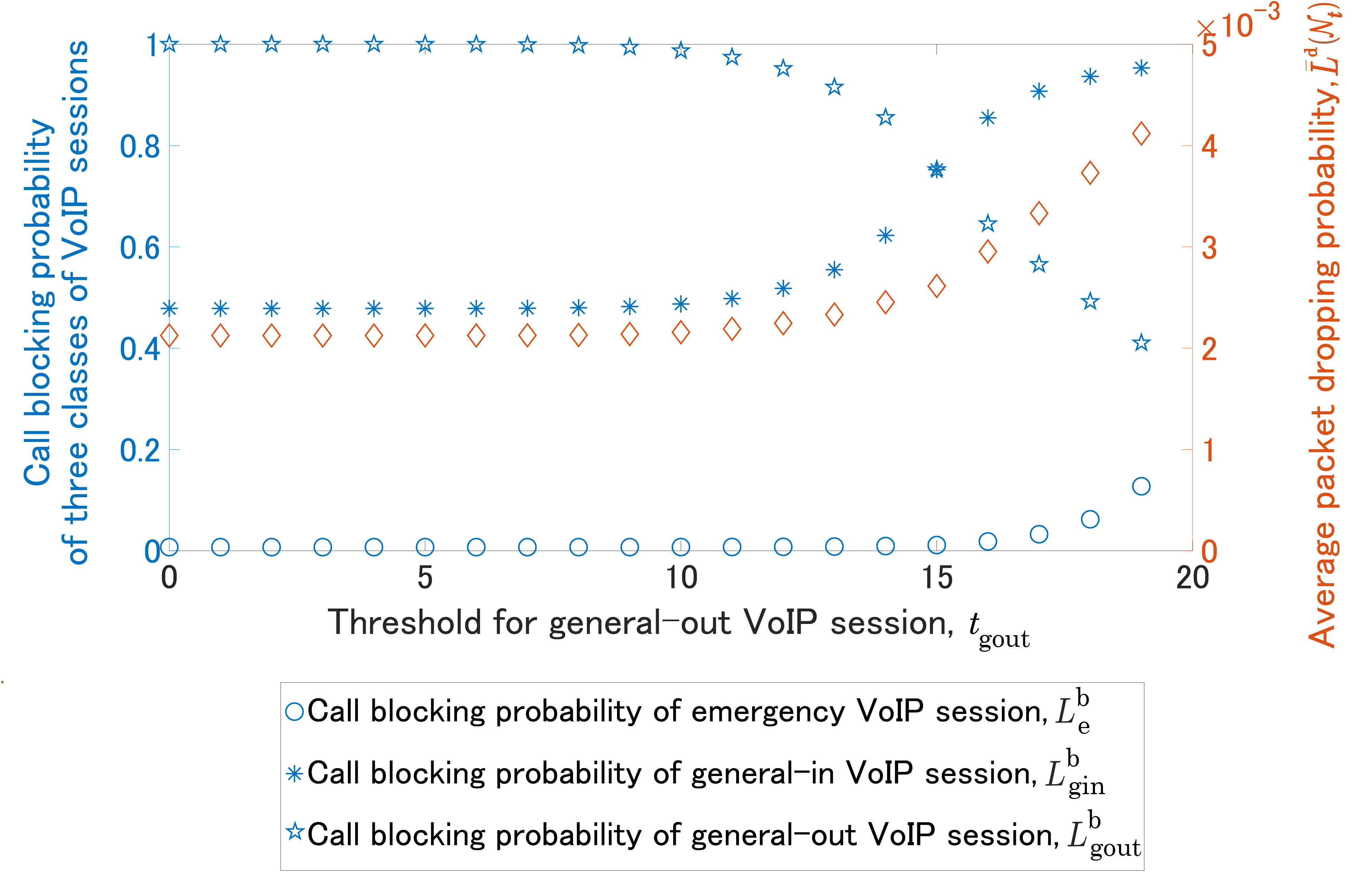}
        \subcaption{When optimal threshold for general-in VoIP session $t^*_{\mathrm{gin}}$ is 15, threshold for general-out VoIP session $t^*_{\mathrm{gout}}$ and each call blocking probability or average packet dropping probability.}
        \label{allcallblock_thgout}
    \end{minipage}
    \caption{Each call blocking probability and average packet dropping probability when varying optimal threshold $(t^*_{\mathrm{gin}},t^*_{\mathrm{gout}})=(15,11)$ when traffic intensity pattern $(\rho_{\mathrm{e}},\rho_{\mathrm{gin}},\rho_{\mathrm{gout}})=(0.45,0.5,0.8)$.}
    \label{allcallblock_e}
\end{figure*}

 Fig. \ref{allcallblock_e} shows the characteristics between the call blocking probability of each VoIP session ($L^{\mathrm{b}}_{\mathrm{e}}, L^{\mathrm{b}}_{\mathrm{gin}}, L^{\mathrm{b }}_{\mathrm{gout}}$) or the average packet dropping probability $\Bar{L}^{\mathrm{d}}({\mathcal N}_{\bm t})$ and the threshold $(t_{\mathrm{gin}}, t_{\mathrm{gout}})$. As shown in Fig. \ref{threshold_e_12}, when the traffic intensity pattern is $(\rho_{\mathrm{e}},\rho_{\mathrm{gin}},\rho_{\mathrm{gout}})=(0.45,0.5,0.8)$, our optimal threshold is $(t^*_{\mathrm{gin}}, t^*_{\mathrm{gout}})=(15,11)$. Therefore, Fig. \ref{allcallblock_thgin} shows the call blocking probability of the three classes of VoIP sessions and $\Bar{L}^{\mathrm{d}}({\mathcal N}_{\bm t})$ when $t_{\mathrm{gin}}$ varies and $t^*_{\mathrm{gout}}=11$. Fig. \ref{allcallblock_thgout} shows the call blocking probability of the three classes of VoIP sessions and  $\Bar{L}^{\mathrm{d}}({\mathcal N}_{\bm t})$ when $t_{\mathrm{gout}}$ varies and $t^*_{\mathrm{gin}}=15$.

 As shown in Figs. \ref{allcallblock_thgin} and  \ref{allcallblock_thgout}, $L^{\mathrm{b}}_{\mathrm{e}}$ increases after both $t_{\mathrm{gin}}$ and $t_{\mathrm{gout}}$ exceed 16. Thus, as shown in Fig. \ref{emergency_block_e}, the reason for  $L^{\mathrm{b}}_{\mathrm{e}}$ to be almost zero is that $t_{\mathrm{gin}}$ and $t_{\mathrm{gout}}$ is lower than 16 in order to reduce the packet dropping probability.
 As shown in Fig. \ref{allcallblock_thgin}, when $t_{\mathrm{gin}}$ is larger than 11, $L^{\mathrm{b}}_{\mathrm{gin}}$ lower than $L^{\mathrm{b}}_{\mathrm{gout}}$. Therefore, when $t_{\mathrm{gin}}$ is larger than 11, the general-in VoIP session has a lower priority than the general-out VoIP session. This means that the priority of the proposed method is not satisfied. When $t_{\mathrm{gout}}$ increases in Fig. \ref{allcallblock_thgout}, the increase in $L^{\mathrm{b}}_{\mathrm{gin}}$ and the decrease in $L^{\mathrm{b}}_{\mathrm{out}}$ are symmetrical. However, when $t_{\mathrm{gin}}$ increases in Fig. \ref{allcallblock_thgin}, the decrease in $L^{\mathrm{b}}_{\mathrm{gin}}$ is higher than the increase in $L^{\mathrm{b}}_{\mathrm{gout}}$. The reason for this is the difference in the effect of each threshold. When $t_{\mathrm{gout}}$ is large, general-out VoIP sessions and general-in VoIP sessions can be accommodated more. However, an increase of $t_{\mathrm{gin}}$ has no effect on the number of accommodated general-out VoIP sessions. Therefore, the reason for the difference in characteristics between Figs. \ref{allcallblock_thgin} and \ref{allcallblock_thgout} is that $t_{\mathrm{gout}}$ is effective for two classes of VoIP sessions, while $t_{\mathrm{gin}}$ is effective only for general-in VoIP sessions. Next, we analyze $(t_{\mathrm{gin}},t_{\mathrm{gout}})$ in total to further determine the characteristics of the packet dropping probability and the thresholds.

\begin{figure}[t]
    \vspace{-20mm}
    \centering
    \includegraphics[clip,width=10.5cm,height=60.0mm]{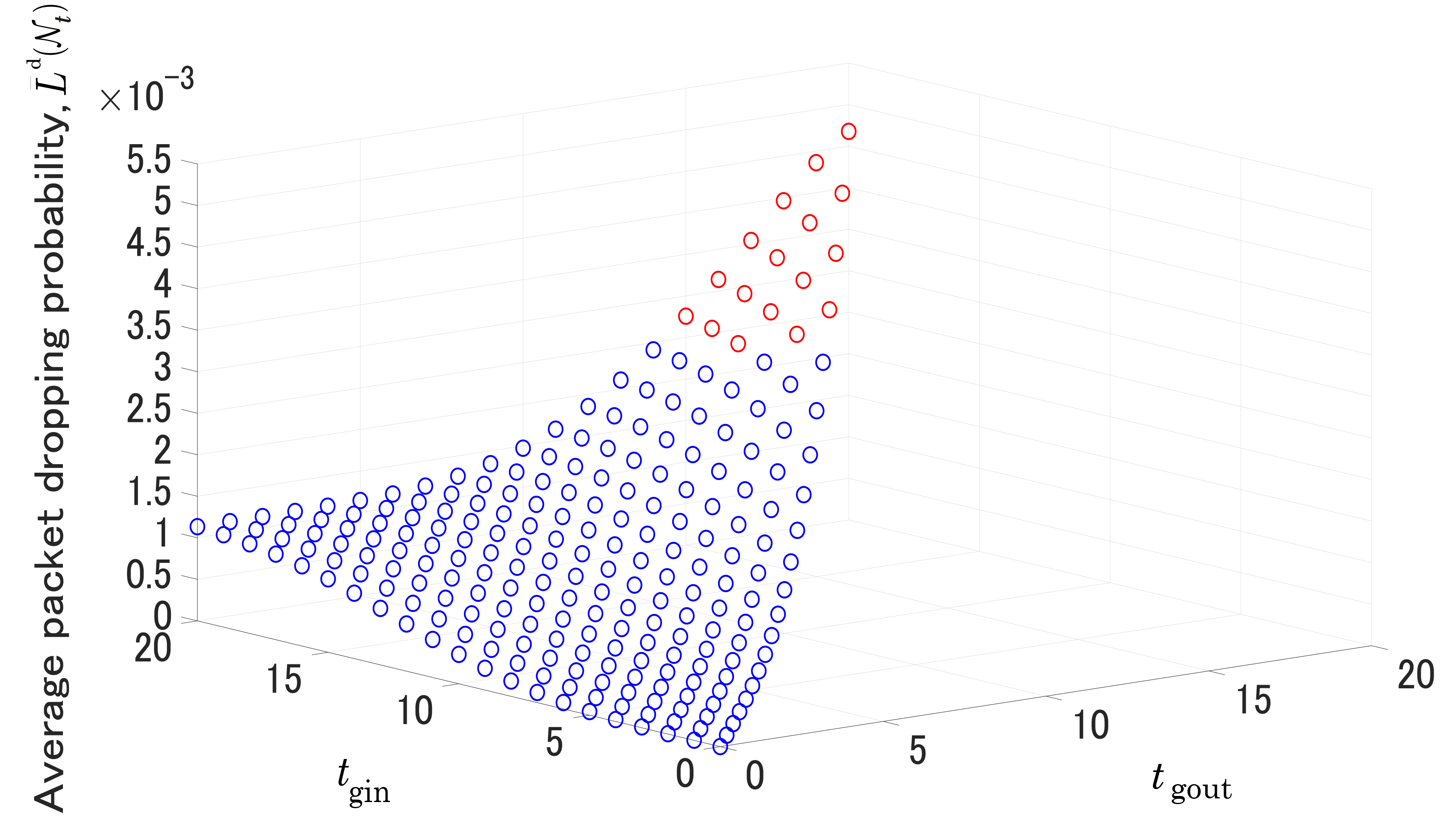}
    \caption{Each threshold and average packet dropping probability when $(\rho_{\mathrm{e}},\rho_{\mathrm{gin}},\rho_{\mathrm{gout}})=(0.05,0.5,0.8)$.}
    \label{packetloss_threshold_0.05}
\end{figure}

\begin{figure}[t]
    \centering
    \hspace{-20mm}
    \includegraphics[clip,width=11.5cm,height=65.0mm]{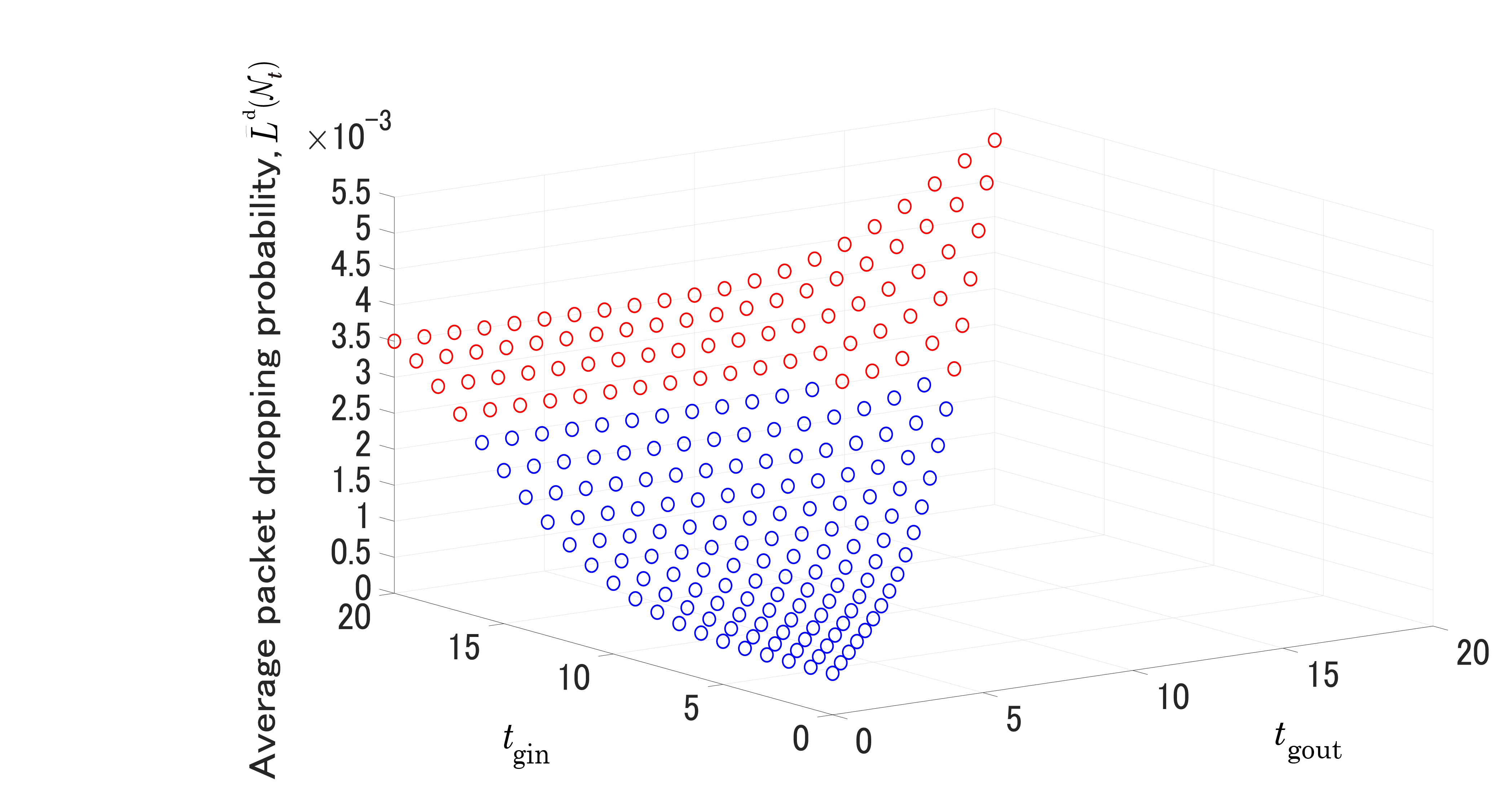}
    \caption{Each threshold and average packet dropping probability when $(\rho_{\mathrm{e}},\rho_{\mathrm{gin}},\rho_{\mathrm{gout}})=(0.45,0.5,0.8)$.}
    \label{packetloss_threshold_0.45}
\end{figure}

\begin{figure}[t]
    \centering
    \includegraphics[clip,width=10cm,height=60.0mm]{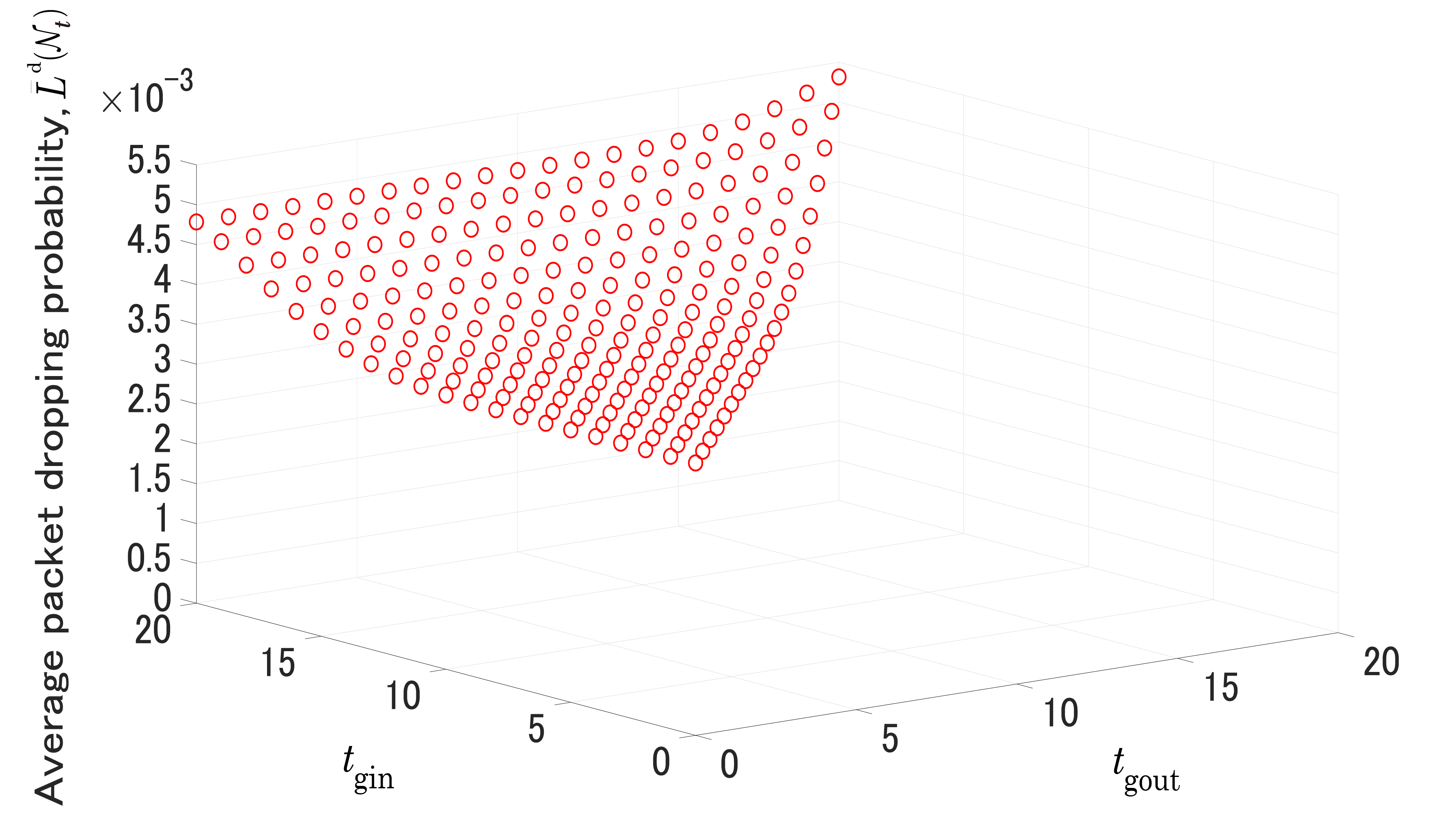}
    \caption{Each threshold and average packet dropping probability when $(\rho_{\mathrm{e}},\rho_{\mathrm{gin}},\rho_{\mathrm{gout}})=(0.95,0.5,0.8)$.}
    \label{packetloss_threshold_0.95}
\end{figure}

Figs. \ref{packetloss_threshold_0.05}--\ref{packetloss_threshold_0.95} shows the characteristics between the average packet dropping probability $\Bar{L}^{\mathrm{d}}({\mathcal N}_{\bm t})$ and the thresholds $(t_{\mathrm{gin}}, t_{\mathrm{gout}})$ when the traffic intensity is $(\rho_{\mathrm{e}},\rho_{\mathrm{gin}},\rho_{\mathrm{gout}})$ = (0.05,0.5,0.8), (0.45,0.5,0.8) and (0.95,0.5,0.8). We set $\Bar{L}^{\mathrm{d}}_{\mathrm{upper}}=$0.0025 \cite{QoS}, which is defined by the Ministry of Internal Affairs and Communications. In Figs. \ref{packetloss_threshold_0.05}--\ref{packetloss_threshold_0.95}, $\Bar{L}^{\mathrm{d}}(\mathcal{N}_{\bm t})\geq \Bar{L}^{\mathrm{d}}_{\mathrm{upper}}$ is plotted in red, and $\Bar{L}^{\mathrm{d}}(\mathcal{N}_{\bm t})< \Bar{L}^{\mathrm{d}}_{\mathrm{upper}}$ is plotted in blue. Moreover, the case of $t_{\mathrm{gin}}<t_{\mathrm{gout}} $ is not plotted due to the requirements of the VoIP session class not being satisfied. As the thresholds $(t_{\mathrm{gin}},{t_\mathrm{gout}})$ increases, $\Bar{L}^{\mathrm{d}}({\mathcal N}_{\bm t})$ also increases. As mentioned in Eq. (\ref{average_packetloss}), $\Bar{L}^{\mathrm{d}}({\mathcal N}_{\bm t})$ is determined by $L^{\mathrm{d}}(\bm{n}_a)$ and $P_{{\bm n}_a}^{\mathrm{session}}$. As the number of accommodated VoIP sessions $\bm{n}_a=(n_{\mathrm{e}}, n_{\mathrm{gin}}, n_{\mathrm{gout}})$ increases, $L^{\mathrm{d}}(\bm {n}_a)$ increases \cite{Narikiyo}. In addition, as $\bm{n}_a$ increases, $P_{{\bm n}_a}^{\mathrm{session}}$ also increases. Thus, when $(t_{\mathrm{gin}}, t_{\mathrm{gout}})$ is large, $L^{\mathrm{d}}(\bm {n}_a)$ increases as larger $\bm{n}_a$ exists in the state space as shown in the Fig. \ref{StateSpace_flow}. In Figs. \ref{packetloss_threshold_0.05}--\ref{packetloss_threshold_0.95}, $\rho_{\mathrm{e}}$ increases in the order of \ref{packetloss_threshold_0.05}, \ref{packetloss_threshold_0.45}, and \ref{packetloss_threshold_0.95}.

\subsection{Analysis of CAC method when varying traffic intensity of general-in VoIP sessions}
\subsubsection{Comparison of conventional CAC method and proposed CAC method for varying traffic intensity of general-in VoIP sessions}

\begin{figure}[t]
    \centering
    \includegraphics[clip,width=11.25cm,height=65.0mm]{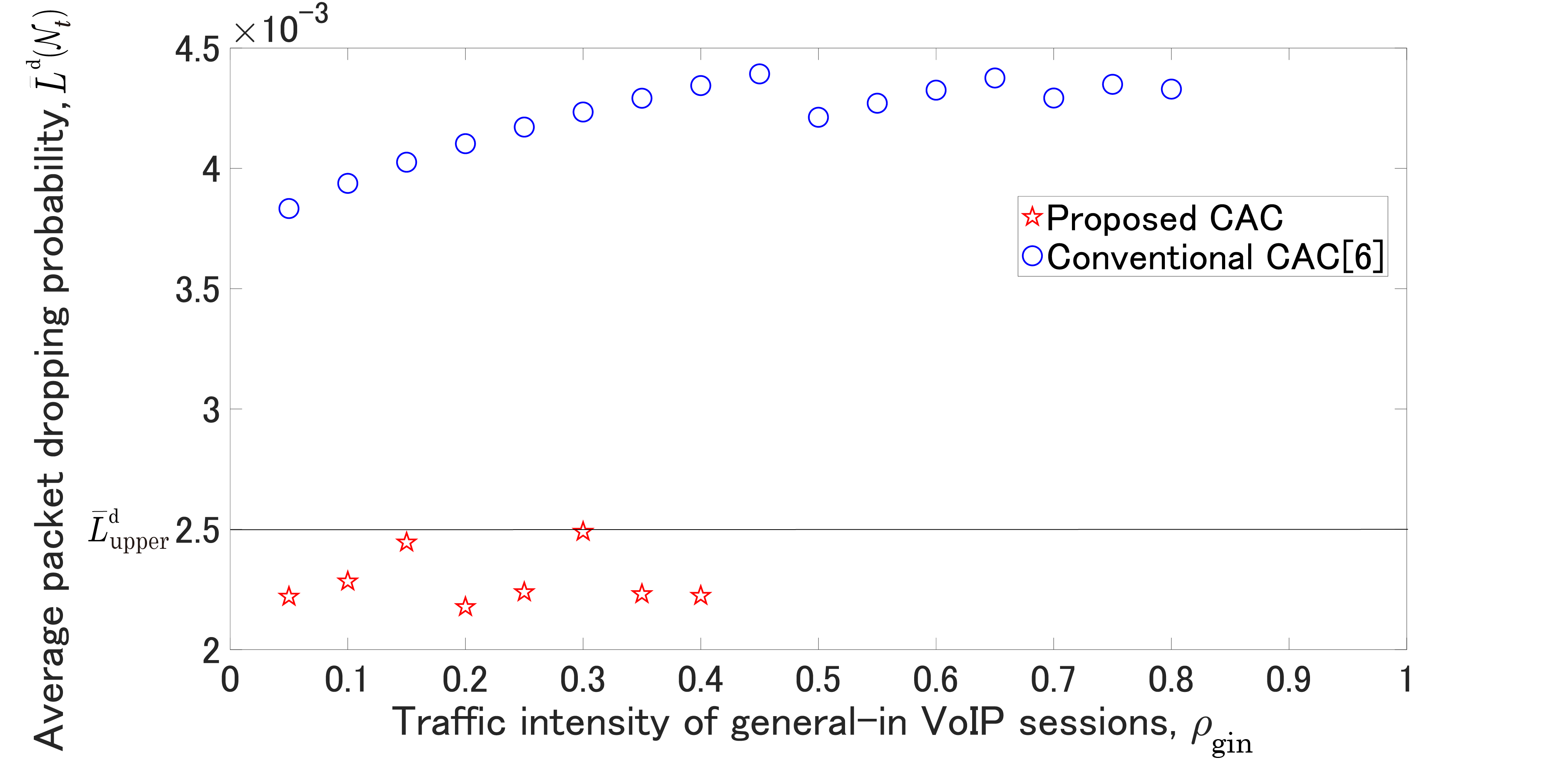}
    \caption{Traffic intensity of general-in VoIP sessions and average packet dropping probability.}
    \label{packetloss_in}
\end{figure}

\begin{figure}[t]
    \centering
    \hspace{-6.5mm}
    \includegraphics[clip,width=11.75cm,height=65.0mm]{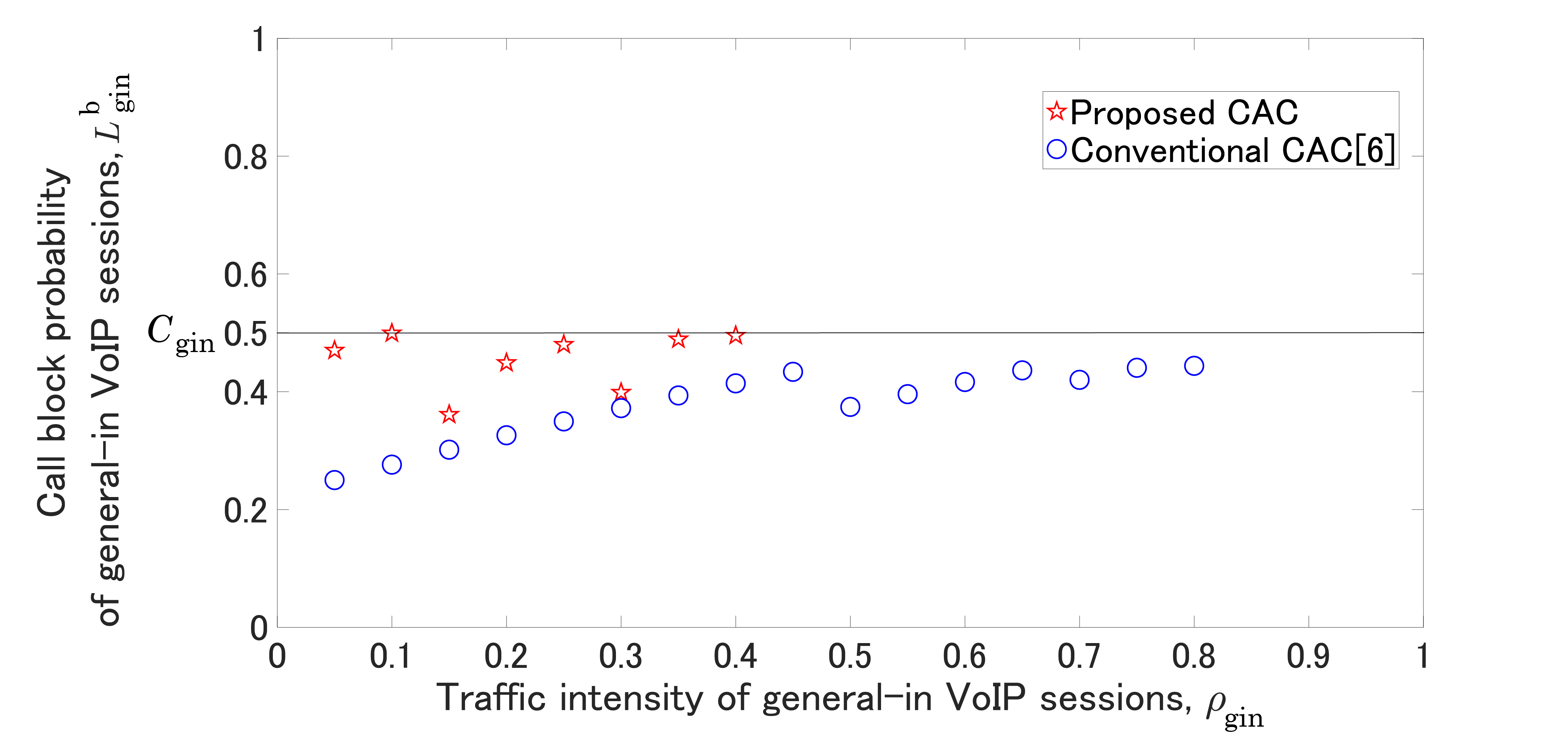}
    \caption{Traffic intensity of general-in VoIP sessions and call blocking probability of general-in VoIP session.}
    \label{in_block_in}
\end{figure}

\begin{figure}[t]
    \centering
    \includegraphics[clip,width=11cm,height=65.0mm]{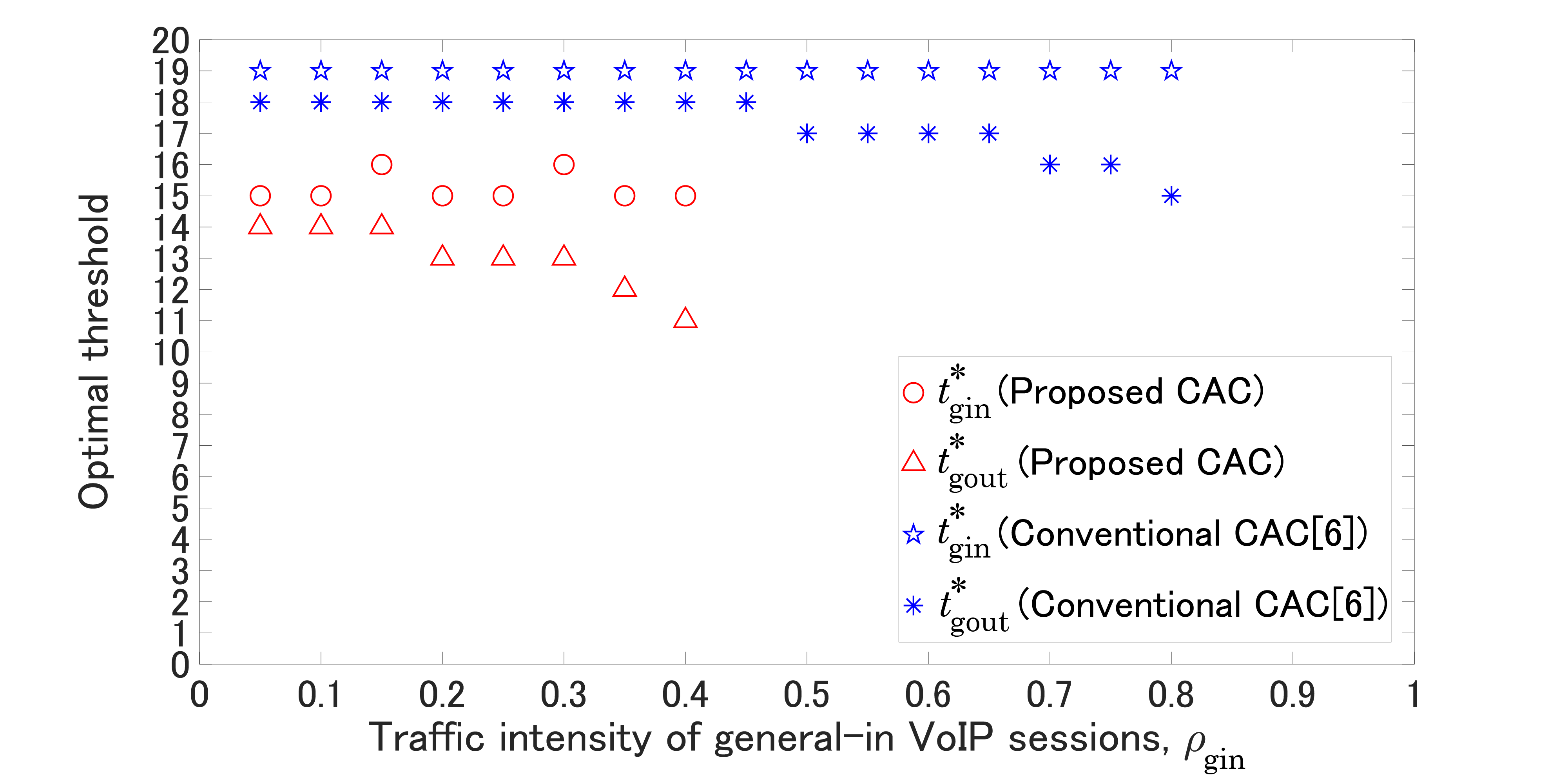}
    \caption{Traffic intensity of general-in VoIP sessions and optimal threshold.}
    \label{threshold_in}
\end{figure}

Figs. \ref{packetloss_in}--\ref{threshold_in} shows the results of the analysis for varying $\rho_{\mathrm{gin}}$. Fig. \ref{packetloss_in} shows the average packet dropping probability $\Bar{L}^{\mathrm{d}}({\mathcal N}_{\bm t})$ with the optimal thresholds $(t^*_{\mathrm{gin}}, t^*_{\mathrm{gout}})$. Fig. \ref{in_block_in} shows the call blocking probability of the general-in VoIP session $L^{\mathrm{b}}_{\mathrm{gin}}$ with the optimal thresholds $(t^*_{\mathrm{gin}}, t^*_{\mathrm{gout}})$. Fig. \ref{threshold_in} shows the optimal thresholds $(t^*_{\mathrm{gin}}, t^*_{\mathrm{gout}})$. Since general-in VoIP sessions are prioritized VoIP sessions as well as emergency VoIP sessions, Fig. \ref{threshold_in} shows similar characteristics to Fig. \ref{threshold_e_12}. For some of the traffic in Fig. \ref{in_block_in}, $L^{\mathrm{b}}_{\mathrm{gin}}$ decreases as $\rho_{\mathrm{gin}}$ increases. This is because the optimal thresholds $(t^*_{\mathrm{gin}}, t^*_{\mathrm{gout}})$ decrease to satisfy the upper bound of the call blocking probability $C_{\mathrm{gin}}$ of the objective function in order to maintain priority between general-in VoIP sessions. In addition, when $t^*_{\mathrm{gout}}$ is small, $L^{\mathrm{b}}_{\mathrm{gin}}$ is lower because more general-in VoIP sessions can be accommodated. These characteristics are the same in Figs. \ref{packetloss_e_12}--\ref{threshold_e_12}.

However, our optimal threshold vector $\bm t^*$ is smaller in Fig. \ref{threshold_in} than in Fig. \ref{threshold_e_12}. This is because $L^{\mathrm{b}}_{\mathrm{gin}}$ changes rapidly as $\rho_{\mathrm{gin}}$ increases or decreases. Thus, we observe that our proposed CAC method is very sensitive to an increase or decrease in $\rho_{\mathrm{gin}}$. This result implies that the setting of $C_{\mathrm{gin}}$ is important for our proposed CAC method. In Sec. 5.4.2), we analyze the call blocking probability characteristics with varying $C_{\mathrm{gin}}$.

\subsubsection{Characteristics analysis with varying upper bound call blocking probability for general-in VoIP sessions}

\begin{figure}[t]
    \centering
    \includegraphics[clip,width=11cm,height=65.0mm]{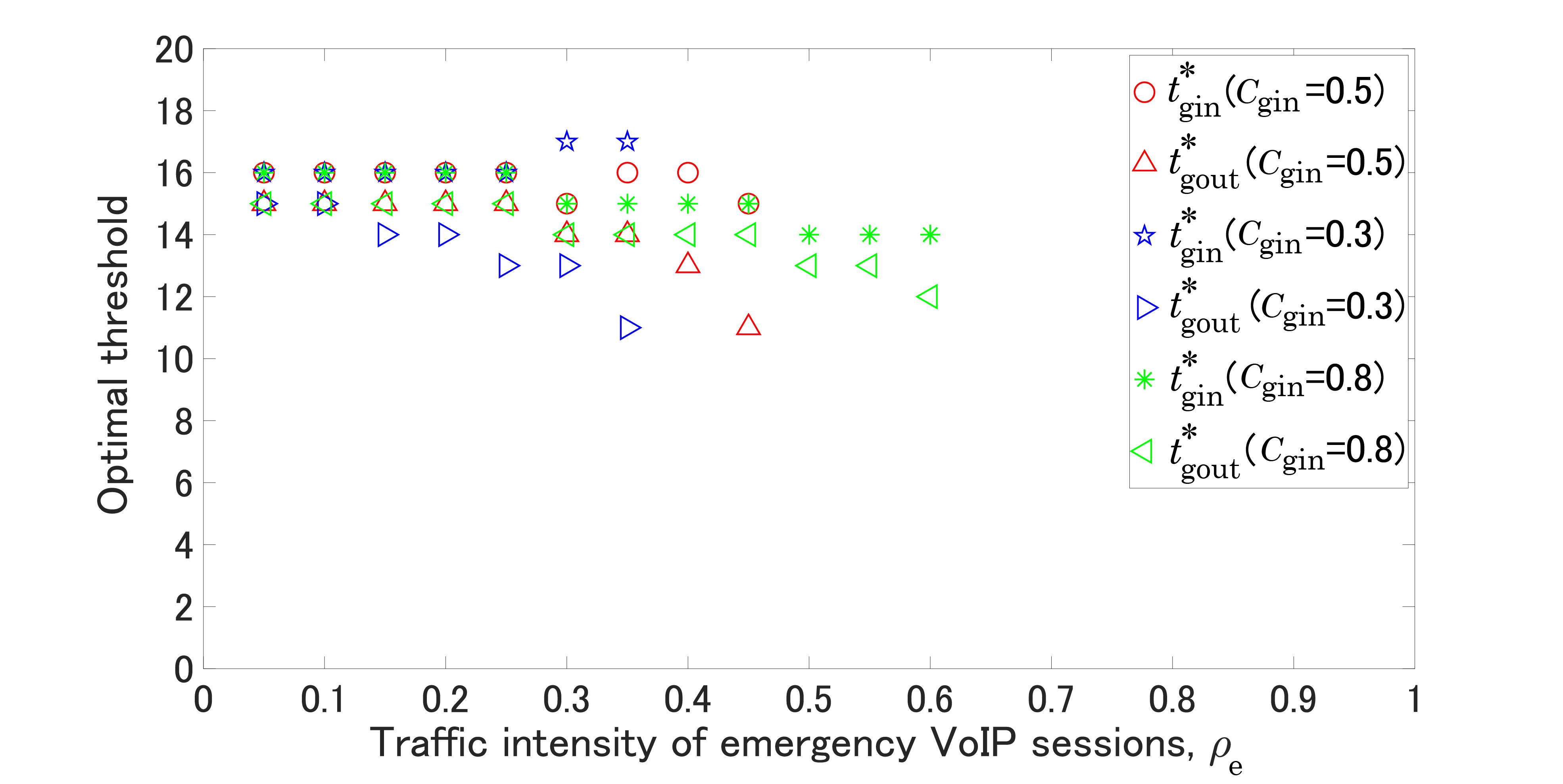}
    \caption{Traffic intensity of emergency VoIP session and optimal threshold when varying upper bound call blocking probability for general-in VoIP session.}
    \label{uppercallblock_gin_e}
\end{figure}

\begin{figure}[t]
    \centering
    \includegraphics[clip,width=11cm,height=65.0mm]{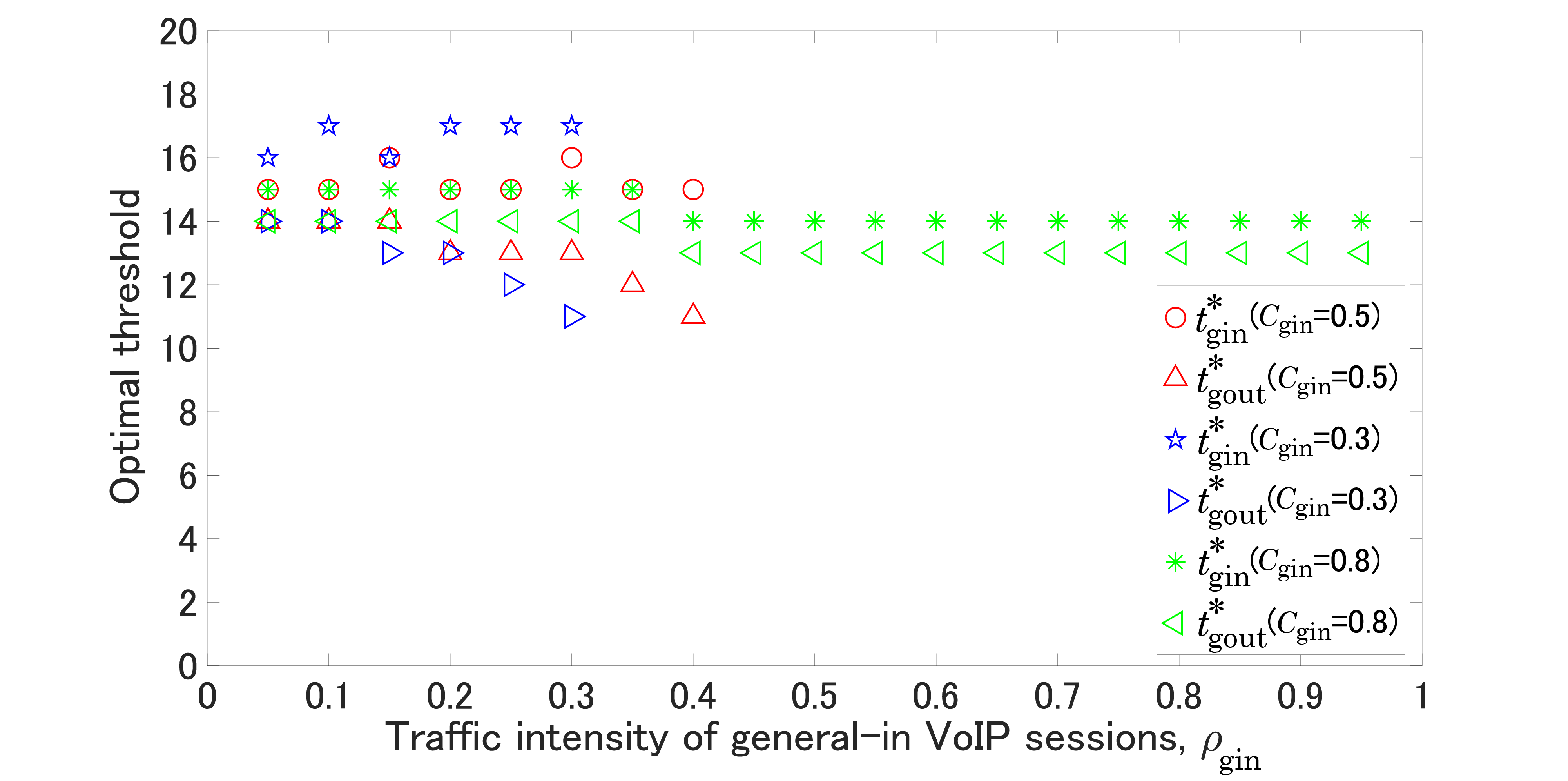}
    \caption{Traffic intensity of general-in VoIP session and optimal threshold when varying upper bound call blocking probability for general-in VoIP session.}
    \label{uppercallblock_gin_gin}
\end{figure}

While $C_{\mathrm{e}}$ is a specified value \cite{uppercallblock}, $C_{\mathrm{gin}}$ can vary depending on the situation. In Figs. \ref{uppercallblock_gin_e} and \ref{uppercallblock_gin_gin}, the traffic intensity pattern is set to $(\rho_{\mathrm{e}},\rho_{\mathrm{gin}},\rho_{\mathrm{gout}})=(0.45,0.5,0.8)$, and $C_{\mathrm{gin}}$ is set to 0.5, 0.3, and 0.8. As shown in Figs. \ref{uppercallblock_gin_e} and \ref{uppercallblock_gin_gin}, when $C_{\mathrm{gin}}$ is set to high, $(t^*_{\mathrm{gin}}, t^*_{\mathrm{gout}})$ can be derived when the traffic intensity of the VoIP session increases. In addition, when $C_{\mathrm{gin}}$ is set to 0.8, the optimal threshold can be derived even though $\rho_{\mathrm{gin}}$ increases compared with $\rho_{\mathrm{e}}$. This is because $C_{\mathrm{gin}}$ is the upper bound for general-in VoIP sessions, and the benefit of setting $C_{\mathrm{gin}}$ high is greatest for general-in VoIP sessions. Therefore, we found that $C_{\mathrm{gin}}$ should be dynamically varied depending on the number of accommodated general-in VoIP sessions $n_{\mathrm{gin}}$ that increase.

\subsection{Analysis of CAC method when varying traffic intensity of general-out VoIP sessions}

\begin{figure}[t]
    \centering
    \includegraphics[clip,width=11cm,height=65.0mm]{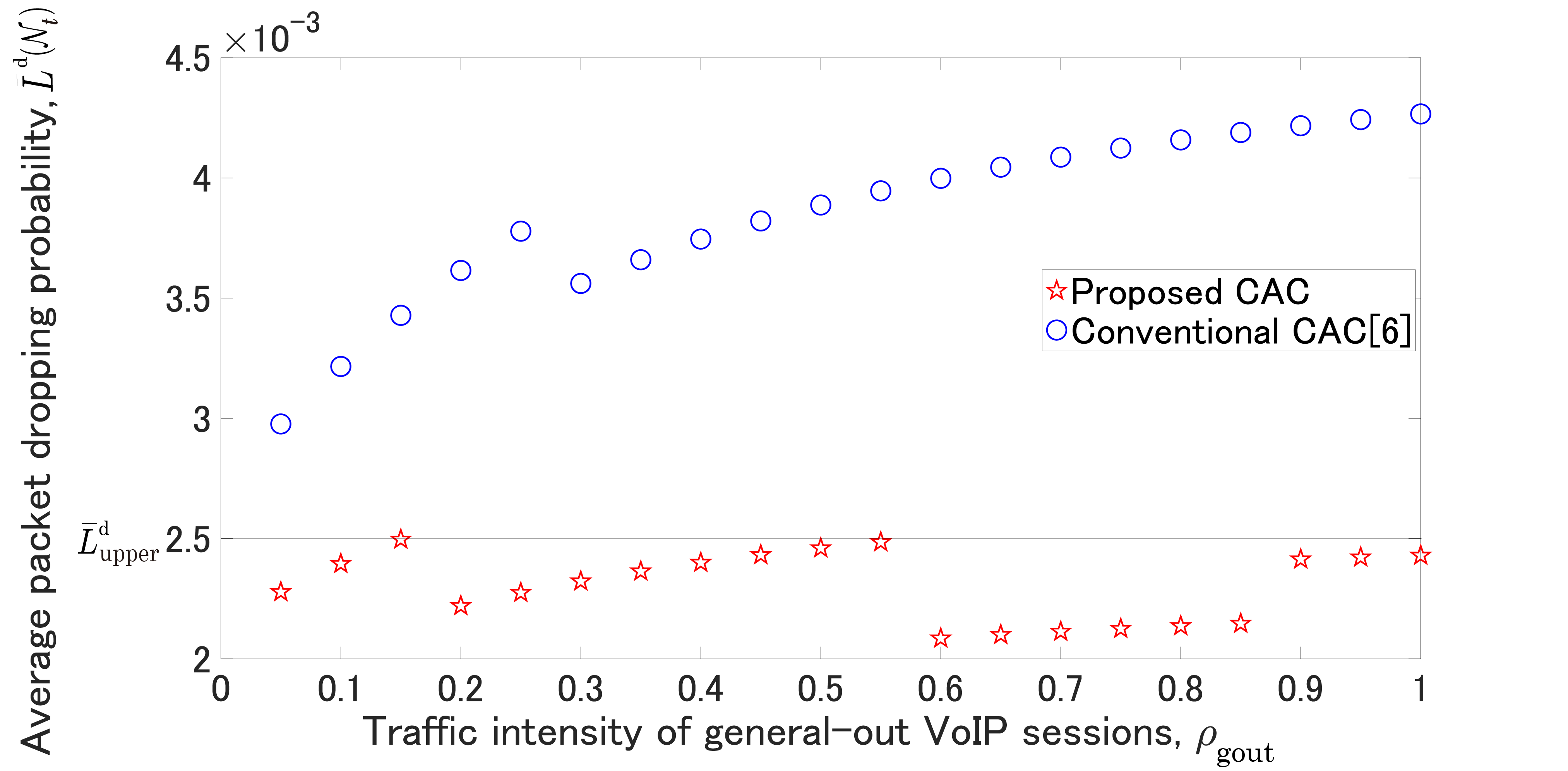}
     \caption{Traffic intensity of general-out VoIP session and average packet dropping probability.}
    \label{packetloss_out}
\end{figure}

\begin{figure}[t]
    \centering
    \includegraphics[clip,width=11cm,height=65.0mm]{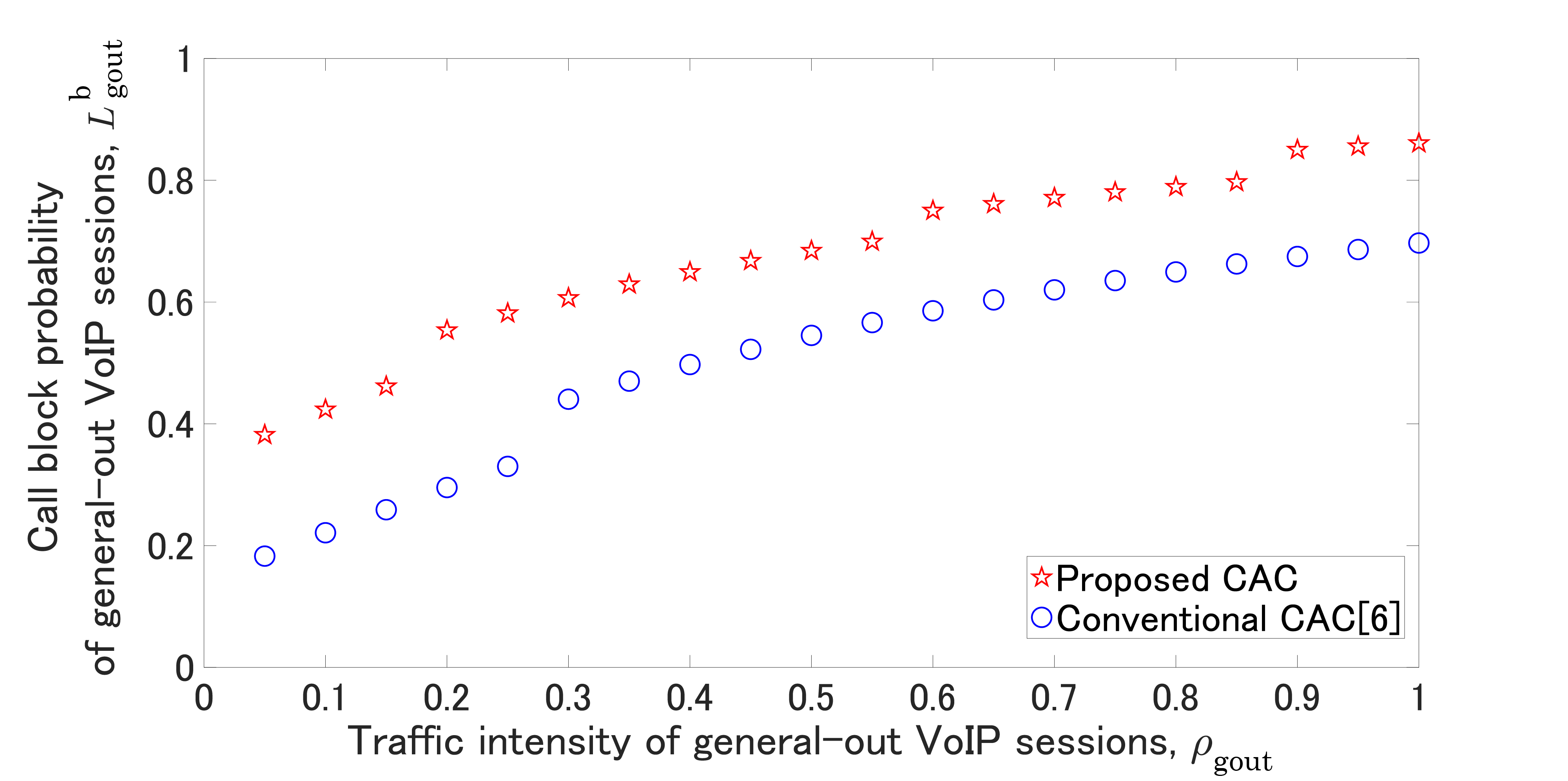}
    \caption{Traffic intensity of general-out VoIP session and call blocking probability of general-out VoIP session.}
    \label{out_block_out}
\end{figure}

\begin{figure}[t]
    \centering
    \includegraphics[clip,width=11cm,height=65.0mm]{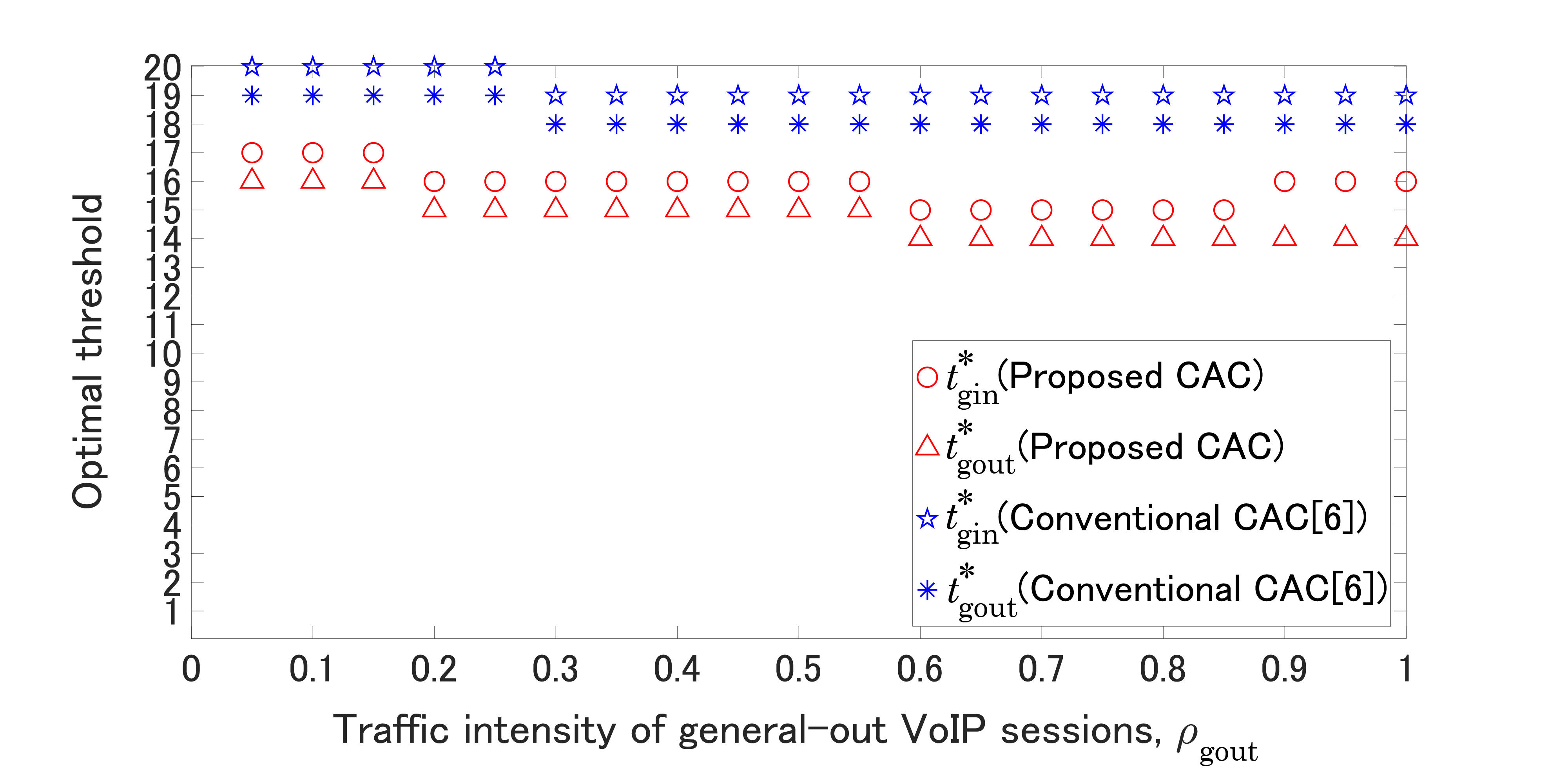}
    \caption{Traffic intensity of general-out VoIP session and optimal threshold.}
    \label{threshold_out}
\end{figure}

Fig. \ref{packetloss_out}--\ref{threshold_out} shows the results of the analysis when $\rho_{\mathrm{gout}}$ is varied. Fig. \ref{packetloss_out} shows the average packet dropping probability $\Bar{L}^{\mathrm{d}}({\mathcal N}_{\bm t})$ at the optimal thresholds $(t^*_{\mathrm{gin}}, t^*_{\mathrm{gout}})$. Fig. \ref{out_block_out} shows the call blocking probability of the general-out VoIP session $L^{\mathrm{b}}_{\mathrm{gout}}$ at the optimal thresholds $(t^*_{\mathrm{gin}}, t^*_{\mathrm{gout}})$. Fig. \ref{threshold_out} shows the optimal thresholds $(t^*_{\mathrm{gin}}, t^*_{\mathrm{gout}})$. As shown in Fig. \ref{threshold_e_12}, $(t^*_{\mathrm{gin}}, t^*_{\mathrm{gout}})$ can not be derived when $\rho_{\mathrm{e}}$ is larger than 0.5. Therefore, we analyze the traffic intensity pattern $(\rho_{\mathrm{e}},\rho_{\mathrm{gin}},\rho_{\mathrm{gout}})=(0.3,0.5,\rho_{\mathrm{out}})$. As shown in Figs. \ref{packetloss_out}, \ref{out_block_out}, and \ref{threshold_out}, $(t^*_{\mathrm{gin}}, t^*_{\mathrm{gout}})$, which controls the packet dropping probability, can be derived even when $\rho_{\mathrm{gout}}$ is varied from 0.05 to 1.0. In addition, $L^{\mathrm{b}}_{\mathrm{gout}}$ in Fig. \ref{out_block_out} does not show the decreasing pattern seen in Figs. \ref{emergency_block_e}, \ref{in_block_in}. The reason for these is that general-out VoIP sessions are the lowest priority VoIP sessions. Thus, we can see that our proposed CAC method is not significantly affected when $\rho_{\mathrm{gout}}$ increases. As a result, we showed that our proposed CAC method could guarantee both communication and connection quality at any traffic intensity.

\vspace{-2mm}
 \section{Conclusion}
We proposed a novel three-dimensional CAC that controls three classes of VoIP sessions and guarantees both communication and connection quality in VoIP networks during emergencies. We evaluated a CAC method considering the packet dropping probability and the call blocking probability. Our optimal threshold could be derived to accommodate the maximum number of general-out VoIP sessions in conditions where the call blocking probability of emergency VoIP sessions and general-in VoIP sessions is under a specified value and where the communication quality is guaranteed. We showed the effectiveness of our proposed method by performing characterization for a variety of traffic conditions.

In future work, we will consider packet delay for a more detailed evaluation of communication quality. In addition, we will consider handover VoIP sessions not newly VoIP sessions. Moreover, our proposed CAC method did not guarantee the connection quality of general-out VoIP sessions by guaranteeing the communication quality. Therefore, we will propose a method with a dedicated call waiting for general-out VoIP sessions.
\acknowledgments
These research results were obtained from the commissioned research (JPJ012368C05601) by National Institute of Information and Communications Technology (NICT), Japan. This work was supported by JSPS KAKENHI Grant Numbers 19K11947, 22K12015, 22H00247.



\end{document}